\documentclass[preprint,showpacs,showkeys]{revtex4}%
\usepackage{amssymb}
\usepackage{amsmath}
\usepackage{graphicx}
\usepackage{epsfig}
\usepackage{bm}
\usepackage{dcolumn}
\usepackage{amsfonts}%
\providecommand{\U}[1]{\protect\rule{.1in}{.1in}}
\begin{document}
\title{Ward Identities in a Two-Dimensional Gravitational Model: Anomalous Amplitude
Revisited Using a Completely Regularization-Independent Mathematical Strategy}
\author{G. Dallabona, P. G. de Oliveira}
\author{O. A. Battistel}

\begin{abstract}
We present a detailed investigation of the anomalous gravitational amplitude
in a simple two-dimensional model with Weyl fermions. We employ a mathematical
strategy that completely avoids any regularization prescription for handling
divergent perturbative amplitudes. This strategy relies solely on the validity
of the linearity of the integration operation and avoids modifying the
amplitudes during intermediate calculations, unlike studies using
regularization methods. Additionally, we adopt arbitrary routings for internal
loop momenta, representing the most general analysis scenario. As expected, we
show that surface terms play a crucial role in both preserving the symmetry
properties of the amplitude and ensuring the mathematical consistency of the
results. Notably, our final perturbative amplitude can be converted into the
form obtained using any specific regularization prescription. We consider
three common scenarios, one of which recovers the traditional results for
gravitational anomalies. However, we demonstrate that this scenario inevitably
breaks the linearity of integration, leading to an undesirable mathematical
situation. This clean and transparent conclusion, enabled by the general
nature of our strategy, would not be apparent in similar studies using
regularization techniques.

\end{abstract}

\maketitle

\address{Departamento de Ciências Exatas, Universidade Federal de Lavras, 37203-022, Lavras, MG, Brazil.}

\address{Departamento de Física, Universidade Federal de Santa Maria, 97119-900,
Santa Maria, RS, Brazil.}

\section{Introduction}

In the late 1960s, the study of neutral electromagnetic pion decay revealed
one of the most remarkable, subtle, and intriguing aspects of quantum field
theory (QFT): the anomaly phenomenon. Specifically, this anomaly is known as
the Adler-Bell-Jackiw (ABJ) anomaly or triangular anomaly, named after the
type of Feynman diagram involved \cite{ADLER, JACKIW1, BARDEEN}. Its
implications extend beyond simply mimicking experimental data; they play a
fundamental role in the structure of QFT itself. For instance, the existence
of three families of six quarks and six leptons as fundamental constituents in
the Standard Model (SM) can be understood as a direct consequence of
anomalies. When anomalous amplitudes are present in a theory, it means that
not all Ward identities associated with different symmetries can be
simultaneously satisfied for those amplitudes. If the broken symmetry is
internal, it leads to internal inconsistencies in the theory, potentially
destroying renormalizability and violating the unitarity of the S-matrix
\cite{KORTHALS}. Consequently, the theory's renormalizability can be rescued
if an anomaly cancellation mechanism exists, where specific combinations of
$1/2$-spin fermions cancel the violations arising from different sectors of
the theory. Such a mechanism, consistent with the SM's structure, necessitates
the existence of six quarks and six leptons \cite{WEINBERG}.

Following the discovery of the ABJ anomaly, numerous other forms of anomalies
have been explored using diverse methods and approaches, both perturbative and
nonperturbative. These include the heat kernel method
\cite{LEUTWYLER1,LEUTWYLER2}, the path integral approach by Fujikawa
\cite{FUJIKAWA1,FUJIKAWA2}, and formal techniques like differential geometry
and cohomology \cite{STORA,LANGOUCHE,ADER,MORITSCH,BONORA,ALVAREZ,RICHTER}.
While these provide elegant formulations, additional tools are often needed to
extract momentum dependence for physical processes. This involves explicitly
evaluating anomalous Feynman diagrams, which are odd tensors (in all
even-dimensional spacetimes) with an odd number of axial-vector vertices and
the remaining vertices being vectors with minimal internal fermionic
propagators. Having at least two Lorentz indices, these tensors cannot
simultaneously preserve all Ward identities (chiral anomalies) while reaching
the expected low-energy limits \cite{SUTHERLAND1,SUTHERLAND2,VELTMAN}. Such
amplitudes share the characteristic that, during dominant-order perturbative
calculations, loop contributions render them divergent. Consequently, their
explicit evaluation relies heavily on the chosen prescription and intermediate
choices, such as internal loop momenta labeling. This presents a dilemma:
either accept the results' dependence on these choices and adjust ambiguous
terms later to achieve desired outcomes, or seek universal procedures for
choice-independent results despite the inherent mathematical ambiguity,
recognizing the anomaly as a fundamental QFT phenomenon.

In fact, a similar universal procedure already exists, proposed in the early
2000s by one of the authors of this work in his doctoral thesis
\cite{ORIMAR-TESE}. This method arose from an effort to develop a
divergence-handling strategy for QFTs that is free from limitations and widely
consistent, allowing tensors and pseudotensors to be treated identically. The
strategy, based on a remarkably simple idea, avoids integrating ill-defined
integrals. Instead, it extracts the physical content by rewriting the
integrand as a sum of finite integrals, surface terms, and purely divergent
objects. Within this framework, divergent quantities lack physical parameters.
Only finite integrals are calculated, while divergent pieces are regrouped
into scalar objects and surface terms. This approach preserves the original
properties of the integrals, enabling broader analysis of relevant physical
processes. This often provides an advantage, allowing sound conclusions in
situations where traditional regularization methods encounter difficulties.
This method is particularly useful when surface terms play a significant role,
as in the case of anomalous perturbative amplitudes, such as the gravitational
anomalies considered in this contribution.

Similar to chiral anomalies in gauge theories, anomalies might arise in the
context of gravitation when fermionic fields couple to the external
gravitational field through the energy-momentum tensor
\cite{CAPPER,DESER,COLEMAN,LANGOUCHE2,LEUTWYLER}. In a seminal work,
Alvarez-Gaum\'{e} and Witten \cite{ALVAREZ-GAUME-2} comprehensively studied
gravitational anomalies in various field theories. They revealed the structure
of these anomalies in higher dimensions and imposed restrictions on theories
compatible with gravity, assuming anomaly cancellation. Specifically,
two-dimensional Weyl spinors exhibit Lorentz and gravitational anomalies
\cite{ALVAREZ-GAUME-2, LANGOUCHE2, HWANG}. More recently, Bertlmann and
Kohlprath \cite{BERTLMANN-1, BERTLMANN-2} employed the dispersion relations
approach in two-dimensional spacetime to investigate Einstein and Weyl
anomalies. They calculated the one-loop Feynman diagram of a Weyl fermion in a
linearized gravitational background, offering a unique perspective on
anomalies compared to ultraviolet regularization methods. Inspired by this
valuable work and the critical nature of the issues raised, we revisit this
intriguing and significant problem in this study. We believe the adopted
procedure can unlock new avenues for analysis. This approach allows us to
obtain results untainted by specific choices typically made during
intermediate calculation steps. In particular, we can clearly examine the role
of arbitrariness associated with internal loop momentum routing in
loop-perturbative amplitudes. It is well-known that shifting the integration
variable for linearly divergent integrals requires compensating with a
corresponding surface term to maintain equality. Therefore, the results for
such amplitudes are expected to depend on chosen internal momentum routings.
Any analysis where these routings are treated as specific combinations of
physical external momenta risks being compromised, as different choices can
lead to different results. This aspect, intimately linked to the role of
surface terms in perturbative calculations, will be demonstrably clarified in
this investigation. Given the absence of these considerations in previous
works and their crucial impact on conclusions, this contribution is warranted.

Building upon the work presented in Ref. \cite{Pedro}, this work offers an alternative 
calculation of the gravitational amplitude described in Bertlmann and Kohlprath's 
studies \cite{BERTLMANN-1,BERTLMANN-2}. We treat the internal loop momenta as arbitrary and avoid
assigning specific values to surface terms during intermediate steps. This
approach directly reveals the structure of ambiguity associated with these
terms and their impact on the qualitative and quantitative interpretation of
results. Surface terms, whose values can vary between methods, are a key
factor in regularization-dependent results. We analyze three commonly
encountered choices associated with different regularization procedures. We
demonstrate that while specific choices allow us to recover traditional
results for gravitational anomalies, these choices inevitably break the
linearity of the integration operation, a fact hidden within traditional
methods. Another notable aspect of our investigation is the connection between
$2D$ gravitational anomalies and the $2D$ chiral anomaly. Our systematic
approach with subamplitudes allows us to identify mathematical structures
shared with simpler theories like $2D$ quantum electrodynamics ($QED_{2}$).
This approach reveals universal aspects of $2D$ anomalies not accessible in
traditional methods.

We organize the work as follows. In Section 2 we establish the theoretical
foundation for our investigation by outlining the expected relationships
among Green's functions (RAGFs) and Ward identities (WIs) associated with
the gravitational amplitude. To facilitate comprehension and simplify the
calculation, we decompose the gravitational amplitude into smaller, manageable
components called subamplitudes. Notably, some of these subamplitudes align
with typical perturbative amplitudes found in simpler QFTs like $QED_{2}$. In
Section 3, we briefly explain the chosen method for handling the divergent
Feynman integrals encountered during the calculation of the gravitational
amplitude. The Section 4 focuses on analyzing the subamplitudes individually.
We calculate each subamplitude and explicitly verify its corresponding RAGFs.
Additionally, a set of conditions required for this purpose is identified.
Leveraging the general results obtained in Section 4, in Section 5 we
investigate the possibility of the gravitational amplitude simultaneously
satisfying its WIs and RAGFs. We emphasize three representative scenarios for
fixing the undefined quantities involved, including the scenario that
generates the usual results for gravitational anomalies. Concluding remarks
and a summary of the key findings are presented in the Section 6.

\section{The Gravitational Amplitude}

In this work we adopt the same model discussed in the Refs. \cite{BERTLMANN-1}
and \cite{BERTLMANN-2} as well as some of definitions and notations stated there.

\subsection{The Model and Definitions}

The background model of our discussions has a Lagrangian whose (linearized)
interaction part may be written as \cite{BONORA}%
\begin{equation}
\mathcal{L}_{I}^{lin}=-\frac{1}{2}h_{\mu\nu}T^{\mu\nu}\ , \label{linearlag}%
\end{equation}
where $T_{\mu\nu}$ is the (symmetric) energy-momentum tensor, explicitly given
by%
\begin{equation}
T^{\mu\nu}=\frac{i}{4}\left[  E_{a}^{\nu}\overline{\psi}\gamma^{a}\left(
\frac{1\pm\gamma^{3}}{2}\right)  \overleftrightarrow{\partial^{\mu}}\psi
+E_{a}^{\mu}\overline{\psi}\gamma^{a}\left(  \frac{1\pm\gamma^{3}}{2}\right)
\overleftrightarrow{\partial^{\nu}}\psi\right]  \ .
\end{equation}
Here $E_{a}^{\mu}$ is the inverse of zweibein $e_{\mu}^{a}$, $\psi$ is the
fermion field, $\gamma^{a}$ are the usual Dirac matrices and $h_{\mu\nu}$ is
the linearized gravitational field, defined through the approximations%
\begin{align*}
g_{\mu\nu}  &  \approx\eta_{\mu\nu}+\kappa h_{\mu\nu}\ ,\ \ \ \ \ \ \ g^{\mu
\nu}\approx\eta^{\mu\nu}-\kappa h^{\mu\nu}\ ,\\
e_{\mu}^{a}  &  \approx\eta_{\mu}^{a}+\frac{1}{2}\kappa h_{\mu}^{a}%
\ ,\ \ \ \ \ \ \ E^{a\mu}\approx\eta^{a\mu}-\frac{1}{2}\kappa h^{a\mu}\ ,
\end{align*}
with $\eta^{\mu\nu}$ being the flat metric. This Lagrangian describe, in two
space-time dimensions, the interaction of a Weyl fermion and a gravitational
background field.

\begin{figure}[h]
\centering
\includegraphics[width=180pt,height=100pt]{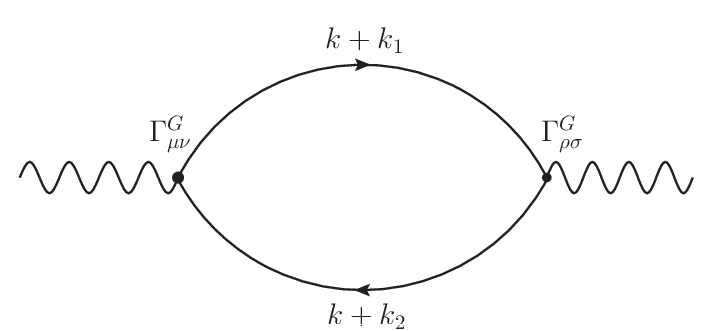} \caption{One-loop
diagrammatic representation for $T_{\mu\nu\sigma\rho}^{G}$.}%
\label{fig_one-loop_grav}%
\end{figure}

The full Green's function which we are interested in is the two-point function%
\begin{equation}
G_{\mu\nu\rho\sigma}\left(  p\right)  =i\int d^{2}x\ e^{ip\cdot x}\left\langle
0\right\vert T\left[  T_{\mu\nu}\left(  x\right)  T_{\rho\sigma}\left(
0\right)  \right]  \left\vert 0\right\rangle \ ,
\end{equation}
which, at one-loop level, is written as%
\begin{equation}
T_{\mu\nu\rho\sigma}^{G}=i\int\frac{d^{2}k}{\left(  2\pi\right)  ^{2}%
}Tr\left\{  \Gamma_{\mu\nu}^{G}\frac{1}{\left[  \not k  +\not k
_{1}-m\right]  }\Gamma_{\rho\sigma}^{G}\frac{1}{\left[  \not k  +\not k
_{2}-m\right]  }\right\}  \ , \label{TG}%
\end{equation}
where
\begin{equation}
\Gamma_{\mu\nu}^{G}=-\frac{i}{4}\left[  \gamma_{\mu}\left(  \left(
k+k_{1}\right)  _{\nu}+\left(  k+k_{2}\right)  _{\nu}\right)  +\gamma_{\nu
}\left(  \left(  k+k_{1}\right)  _{\mu}+\left(  k+k_{2}\right)  _{\mu}\right)
\right]  \frac{\left(  1\pm\gamma_{3}\right)  }{2}\ ,
\end{equation}
gives the Feynman rule for the vertex function. A diagrammatic representation
of $T_{\mu\nu\rho\sigma}^{G}$ can be seen in the Fig. (\ref{fig_one-loop_grav}%
). 

Observe that we have adopted general labels for the internal propagators,
namely $k+k_{1}$ and $k+k_{2}$, and, for convenience, taken the fermion as
being massive. Given this routing, the external momentum is identified as
$p=k_{2}-k_{1}$. The adoption of arbitrary labels is an important attitude in
perturbative calculations in general but is of special importance in the
presently considered problem.\emph{ }Once the power counting of the loop
momentum point out divergence degree higher than the logarithmic one, it is
expected that the result is dependent on the routing adopted for the internal
lines momenta. The arbitrary choice guarantee that such dependence can be
identified in the final results. If the internal momenta are label in terms of
external momenta such that the sum $k_{1}+k_{2}$ is not zero, terms which
would be nonphysical will be mixed with label independent terms, compromising
then the analysis.

The explicit calculation of (\ref{TG}), especially with general labels for the
internal propagators, is very long and tedious. However, the conclusions
extracted from are strongly connected with the calculation details. In order
to make an useful investigation, we need to adopt a systematic way to present
such calculations. Having this in mind, in the present work, we adopt, for the
sake of clarity, a particular systematization. We split out $T_{\mu\nu
\rho\sigma}^{G}$ into three sets of amplitudes as%
\begin{equation}
T_{\mu\nu\rho\sigma}^{G}=-\frac{i}{64}\left\{  T_{\mu\nu\rho\sigma}^{\left(
V\right)  }+T_{\mu\nu\rho\sigma}^{\left(  AV\right)  }+T_{\mu\nu\rho\sigma
}^{\left(  A\right)  }\right\}  \ ,
\end{equation}
where each set is composed by a sum of two-point subamplitudes, namely%
\begin{align}
T_{\mu\nu\rho\sigma}^{\left(  V\right)  }  &  =4\left[  T_{\nu\sigma;\mu\rho
}^{VV}\right]  +2p_{\sigma}\left[  T_{\nu;\mu\rho}^{VV}\right]  +2p_{\nu
}\left[  T_{\sigma;\mu\rho}^{VV}\right]  +p_{\nu}p_{\sigma}\left[  T_{\mu\rho
}^{VV}\right] \nonumber\\
&  +\left(  \mu\longleftrightarrow\nu\right)  +\left(  \sigma
\longleftrightarrow\rho\right)  \ , \label{V_Sector}%
\end{align}%
\begin{align}
\pm T_{\mu\nu\rho\sigma}^{\left(  AV\right)  }  &  =4\left[  T_{\nu\sigma
;\mu\rho}^{AV}\right]  +2p_{\sigma}\left[  T_{\nu;\mu\rho}^{AV}\right]
+2p_{\nu}\left[  T_{\sigma;\mu\rho}^{AV}\right]  +p_{\nu}p_{\sigma}\left[
T_{\mu\rho}^{AV}\right] \nonumber\\
&  +4\left[  T_{\nu\sigma;\mu\rho}^{VA}\right]  +2p_{\sigma}\left[  T_{\nu
;\mu\rho}^{VA}\right]  +2p_{\nu}\left[  T_{\sigma;\mu\rho}^{VA}\right]
+p_{\nu}p_{\sigma}\left[  T_{\mu\rho}^{VA}\right] \nonumber\\
&  +\left(  \mu\longleftrightarrow\nu\right)  +\left(  \sigma
\longleftrightarrow\rho\right)  \ , \label{AV_Sector}%
\end{align}%
\begin{align}
T_{\mu\nu\rho\sigma}^{\left(  A\right)  }  &  =4\left[  T_{\nu\sigma;\mu\rho
}^{AA}\right]  +2p_{\sigma}\left[  T_{\nu;\mu\rho}^{AA}\right]  +2p_{\nu
}\left[  T_{\sigma;\mu\rho}^{AA}\right]  +p_{\nu}p_{\sigma}\left[  T_{\mu\rho
}^{AA}\right] \nonumber\\
&  +\left(  \mu\longleftrightarrow\nu\right)  +\left(  \sigma
\longleftrightarrow\rho\right)  \ . \label{A_Sector}%
\end{align}
The subamplitudes appearing in the above expressions are defined by%
\begin{equation}
T_{\rho\sigma}^{ij}=\int\frac{d^{2}k}{\left(  2\pi\right)  ^{2}}Tr\left\{
\left[  \Gamma_{i}\right]  _{\rho}\frac{1}{\not k  +\not k  _{1}-m}\left[
\Gamma_{j}\right]  _{\sigma}\frac{1}{\not k  +\not k  _{2}-m}\right\}  ~,
\label{T2}%
\end{equation}%
\begin{equation}
T_{\mu;\rho\sigma}^{ij}=\int\frac{d^{2}k}{\left(  2\pi\right)  ^{2}}\left(
k+k_{1}\right)  _{\mu}Tr\left\{  \left[  \Gamma_{i}\right]  _{\rho}\frac
{1}{\not k  +\not k  _{1}-m}\left[  \Gamma_{j}\right]  _{\sigma}\frac
{1}{\not k  +\not k  _{2}-m}\right\}  \ , \label{T3}%
\end{equation}

\begin{equation}
T_{\mu\nu;\rho\sigma}^{ij}=\int\frac{d^{2}k}{\left(  2\pi\right)  ^{2}}\left(
k+k_{1}\right)  _{\mu}\left(  k+k_{1}\right)  _{\nu}Tr\left\{  \left[
\Gamma_{i}\right]  _{\rho}\frac{1}{\not k  +\not k  _{1}-m}\left[  \Gamma
_{j}\right]  _{\sigma}\frac{1}{\not k  +\not k  _{2}-m}\right\}  \ .
\label{T4}%
\end{equation}
In the expressions above the quantities $\Gamma_{i}$ are vertex operators
belonging to the set $\Gamma_{i}=\left\{  \Gamma_{S},\Gamma_{P},\Gamma
_{V},\Gamma_{A}\right\}  =\{1,\gamma_{3},\gamma_{\alpha},\gamma_{\alpha}%
\gamma_{3}\}$. In addition to the calculation aspect, such introduced
systematization will help us to verify the consistency of the obtained results
in a wider sense. Note that the first set of subamplitudes appears in
renormalizable theories like the $QED_{2}$ \cite{SCHWINGER, BERTLMANN-BOOK,
ADAM}. This allow us to add an additional aspect to the investigation.

\subsection{ Relations among Green's functions and Ward Identities}

Along the difficult task of constructing a consistent interpretation of the
perturbative amplitudes in QFT's, when the involved quantities are divergent,
a special recourse can play a very important role. We denominated it as
relations among Green's functions. Such relations can be stated always we have
a Lorentz index attached to a perturbative amplitude. They are constructed by
using simple ingredients like the Dirac algebra, cyclicity and linearity of
the trace operation and, especially, the linearity of the integration
operation. In particular, preservation of the linearity in the integration
operation involving divergent Feynman integrals is not a trivial job, as we
will see along this work.

Let us consider, in this section, the relevant RAGFs for all required
subamplitudes of (\ref{TG}). In fact, to state the referred relations is a
trivial task. In order to exemplify the procedure, we consider the algebraic
identity
\begin{align}
&  \left(  {k_{2}}-k_{1}\right)  ^{\nu}\left\{  \gamma_{\mu}\frac{1}{\left[
\left(  \not k  +\not k  _{1}\right)  -m\right]  }\gamma_{\nu}\frac{1}{\left[
\left(  \not k  +\not k  _{2}\right)  -m\right]  }\right\} \nonumber\\
&  =\left\{  \gamma_{\mu}\frac{1}{\left[  \left(  \not k  +\not k
_{1}\right)  -m\right]  }\right\}  -\left\{  \gamma_{\mu}\frac{1}{\left[
\left(  \not k  +\not k  _{2}\right)  -m\right]  }\right\}  \ ,
\end{align}
which is obtained through the ingredients cited above. In practical terms,
through this operation is possible to cancel out an internal propagator. When
the integration in the loop momentum $k$ is taken, after taken the traces in
both sides, this algebraic identity will be converted into a genuine RAGFs
involving the contraction of the polarization tensor $T_{\mu\nu}^{VV}$ with
the external momentum $p^{\nu}$ and two one-point vector amplitudes defined by%
\begin{equation}
T_{\mu}^{V}\left(  k_{i}\right)  =\int\frac{d^{2}k}{\left(  2\pi\right)  ^{2}%
}Tr\left\{  \gamma_{\mu}\frac{1}{\left[  \left(  \not k  +\not k  _{i}\right)
-m\right]  }\right\}  \ .
\end{equation}
Explicitly, we get the following RAGFs%
\begin{equation}
p^{\sigma}\left[  T_{\sigma\rho}^{VV}\left(  k_{1},k_{2}\right)  \right]
=T_{\rho}^{V}\left(  k_{1}\right)  -T_{\rho}^{V}\left(  k_{2}\right)  \ ,
\label{VV_RAGF_1}%
\end{equation}
and, in a similar way,%
\begin{equation}
p^{\rho}\left[  T_{\sigma\rho}^{VV}\left(  k_{1},k_{2}\right)  \right]
=T_{\sigma}^{V}\left(  k_{1}\right)  -T_{\sigma}^{V}\left(  k_{2}\right)  \ .
\label{VV_RAGF_2}%
\end{equation}
If the $T_{\sigma\rho}^{VV}$ and $T_{\rho}^{V}$ amplitudes are evaluated,
through some particular procedure, in such a way that the final results are in
disagreement with the RAGFs, it means, undoubtedly, that the linearity in the
integration operation was violated through the operations made. Of course,
this is not the adequate situation if one want to make predictions in
perturbative treatments of a model or theory. In this sense, the RAGFs give us
a powerful test of consistency of a method used to calculate divergent
perturbative amplitudes.

Following this procedure it is possible to state two relevant RAGFs for
$T_{\mu\nu\rho\sigma}^{G}$ through the contractions of (\ref{TG}) with the
momentum $p^{\mu}$ and the metric $g^{\mu\nu}$.\ According to our previously
introduced notation, we can write%
\begin{align}
p^{\mu}T_{\mu\nu\rho\sigma}^{G}  &  =-\frac{i}{64}\left\{  \left[  p^{\mu
}T_{\mu\nu\rho\sigma}^{\left(  V\right)  }\right]  +\left[  p^{\mu}T_{\mu
\nu\rho\sigma}^{\left(  AV\right)  }\right]  +\left[  p^{\mu}T_{\mu\nu
\rho\sigma}^{\left(  A\right)  }\right]  \right\}  \ ,\\
g^{\mu\nu}T_{\mu\nu\rho\sigma}^{G}  &  =-\frac{i}{64}\left\{  \left[
g^{\mu\nu}T_{\mu\nu\rho\sigma}^{\left(  V\right)  }\right]  +\left[  g^{\mu
\nu}T_{\mu\nu\rho\sigma}^{\left(  AV\right)  }\right]  +\left[  g^{\mu\nu
}T_{\mu\nu\rho\sigma}^{\left(  A\right)  }\right]  \right\}  \ ,
\end{align}
which, in terms of the subamplitudes, means%
\begin{align}
p^{\mu}\left[  T_{\mu\nu\rho\sigma}^{\left(  V\right)  }\right]   &
=4\left\{  \left[  p^{\mu}T_{\mu\sigma;\nu\rho}^{VV}\right]  +\left[  p^{\mu
}T_{\nu\sigma;\mu\rho}^{VV}\right]  \right\} \nonumber\\
&  +2p_{\sigma}\left\{  \left[  p^{\mu}T_{\nu;\mu\rho}^{VV}\right]  +\left[
p^{\mu}T_{\mu;\nu\rho}^{VV}\right]  \right\} \nonumber\\
&  +2p_{\nu}\left[  p^{\mu}T_{\sigma;\mu\rho}^{VV}\right]  +p_{\nu}p_{\sigma
}\left[  p^{\mu}T_{\mu\rho}^{VV}\right] \nonumber\\
&  +p^{2}\left\{  p_{\sigma}\left[  T_{\nu\rho}^{VV}\right]  +2\left[
T_{\sigma;\nu\rho}^{VV}\right]  \right\}  +\left(  \sigma\leftrightarrow
\rho\right)  \ ,
\end{align}%
\begin{align}
p^{\mu}\left[  T_{\mu\nu\rho\sigma}^{\left(  A\right)  }\right]   &
=4\left\{  \left[  p^{\mu}T_{\nu\sigma;\mu\rho}^{AA}\right]  +\left[  p^{\mu
}T_{\mu\sigma;\nu\rho}^{AA}\right]  \right\} \nonumber\\
&  +2p_{\sigma}\left\{  \left[  p^{\mu}T_{\nu;\mu\rho}^{AA}\right]  +\left[
p^{\mu}T_{\mu;\nu\rho}^{AA}\right]  \right\} \nonumber\\
&  +2p_{\nu}\left[  p^{\mu}T_{\sigma;\mu\rho}^{AA}\right]  +p_{\nu}p_{\sigma
}\left[  p^{\mu}T_{\mu\rho}^{AA}\right] \nonumber\\
&  +p^{2}\left\{  p_{\sigma}\left[  T_{\nu\rho}^{AA}\right]  -2\left[
T_{\sigma;\nu\rho}^{AA}\right]  \right\}  +\left(  \sigma\leftrightarrow
\rho\right)  \ ,
\end{align}%
\begin{align}
p^{\mu}\left[  T_{\mu\nu\rho\sigma}^{\left(  AV\right)  }\right]   &
=\pm4\left\{  \left[  p^{\mu}T_{\nu\sigma;\mu\rho}^{VA}+p^{\mu}T_{\nu
\sigma;\mu\rho}^{AV}\right]  +\left[  p^{\mu}T_{\mu\sigma;\nu\rho}^{VA}%
+p^{\mu}T_{\mu\sigma;\nu\rho}^{AV}\right]  \right\} \nonumber\\
&  \pm2p_{\sigma}\left\{  \left[  p^{\mu}T_{\nu;\mu\rho}^{VA}+p^{\mu}%
T_{\nu;\mu\rho}^{AV}\right]  +\left[  p^{\mu}T_{\mu;\nu\rho}^{VA}+p^{\mu
}T_{\mu;\nu\rho}^{AV}\right]  \right\} \nonumber\\
&  \pm2p_{\nu}\left[  p^{\mu}T_{\sigma;\mu\rho}^{VA}+p^{\mu}T_{\sigma;\mu\rho
}^{AV}\right]  \pm p_{\nu}p_{\sigma}\left[  p^{\mu}T_{\mu\rho}^{VA}+p^{\mu
}T_{\mu\rho}^{AV}\right] \nonumber\\
&  \pm p^{2}\left\{  p_{\sigma}\left[  T_{\nu\rho}^{VA}+T_{\nu\rho}%
^{AV}\right]  +2\left[  T_{\sigma;\nu\rho}^{VA}+T_{\sigma;\nu\rho}%
^{AV}\right]  \right\}  +\left(  \sigma\leftrightarrow\rho\right)  \ ,
\end{align}
as well as%
\begin{align}
g^{\mu\nu}\left[  T_{\mu\nu\rho\sigma}^{\left(  V\right)  }\right]   &
=8\left[  g^{\mu\nu}T_{\mu\sigma;\nu\rho}^{VV}\right]  +4p^{\sigma}\left[
g^{\mu\nu}T_{\mu;\nu\rho}^{VV}\right] \nonumber\\
&  +4\left[  p^{\mu}T_{\sigma;\mu\rho}^{VV}\right]  +2p_{\sigma}\left[
p^{\mu}T_{\mu\rho}^{VV}\right]  +\left(  \sigma\leftrightarrow\rho\right)  \ ,
\end{align}%
\begin{align}
g^{\mu\nu}\left[  T_{\mu\nu\rho\sigma}^{\left(  A\right)  }\right]   &
=8\left[  g^{\mu\nu}T_{\mu\sigma;\nu\rho}^{AA}\right]  +4p^{\sigma}\left[
g^{\mu\nu}T_{\mu;\nu\rho}^{AA}\right] \nonumber\\
&  +4\left[  p^{\mu}T_{\sigma;\mu\rho}^{AA}\right]  +2p_{\sigma}\left[
p^{\mu}T_{\mu\rho}^{AA}\right]  +\left(  \sigma\leftrightarrow\rho\right)  \ ,
\end{align}
and%
\begin{align}
g^{\mu\nu}\left[  T_{\mu\nu\rho\sigma}^{\left(  AV\right)  }\right]   &
=\pm8\left[  g^{\mu\nu}T_{\nu\sigma;\mu\rho}^{VA}+g^{\mu\nu}T_{\nu\sigma
;\mu\rho}^{AV}\right] \nonumber\\
&  \pm4p_{\sigma}\left[  g^{\mu\nu}T_{\nu;\mu\rho}^{VA}+g^{\mu\nu}T_{\nu
;\mu\rho}^{AV}\right] \nonumber\\
&  \pm4\left[  p^{\mu}T_{\sigma;\mu\rho}^{VA}+p^{\mu}T_{\sigma;\mu\rho}%
^{AV}\right] \nonumber\\
&  \pm2p_{\sigma}\left[  p^{\mu}T_{\mu\rho}^{VA}+p^{\mu}T_{\mu\rho}%
^{AV}\right]  +\left(  \sigma\leftrightarrow\rho\right)  \text{ }.
\end{align}
In practical terms, we need to state the RAGFs for all subamplitudes defined
in Eqs. (\ref{T2}), (\ref{T3}), and (\ref{T4}). Since the procedure to obtain
such RAGFs presents no difficulties, we just list all of them in the appendix
(\ref{Ap_GFRs}).

At this point it is interesting to see that the RAGFs are, strictly speaking,
mathematical identities which are valid in a way independent of the particular
context. Then, one would not expected that they were violated by any
calculation procedure. On the other hand, one would not expect that the Ward
identities must be satisfied automatically in the perturbative calculations
due to the fact that they are stated by assuming translational invariance as
an ingredient. Such property is not contained in the amplitudes constructed
through the Feynman rules since, in cases where the divergence degree involved
is higher than the logarithmic one, the result is dependent in the routing
assumed for the loops internal lines momenta. Two distinct labels obeying
energy-momentum conservation in all vertexes will generate results which can
differ by terms that are proportional to surface terms. The coefficients of
such terms are ambiguous combination of the internal lines momenta. In this
way, the WIs are expected to be broken in situations where surface terms are
involved. Within this context, it is the preservation that must be considered
as a special accident and not the violation. So, we must, before considering
the content of an explicit mathematical form of an amplitude, verify if the
adopted procedure does not breaks the pertinent RAGFs. In order to satisfy the
WIs a new ingredient must be added to the implication of Feynman rules. The
usual one is the adoption of a regularization procedure. In the present
investigation we adopt a procedure which does not modify the amplitudes in the
intermediary steps of the calculations. The final form is a pure implication
of the Feynman rules such that, from the results, the ones corresponding to
other prescriptions can be obtained.

Given the symmetries of the considered model \cite{BERTLMANN-1}, the
energy-momentum tensor $T_{\mu\nu}$ is expected to has the following three
properties,%
\begin{equation}
T_{\mu\nu}=T_{\nu\mu},\ \ \nabla^{\mu}T_{\mu\nu}=0,\ \ g^{\mu\nu}T_{\mu\nu
}=0\ ,
\end{equation}
which imply, respectively, three canonical WIs (for massless fermions)%
\begin{equation}
\left\{
\begin{array}
[c]{c}%
T_{\mu\nu\rho\sigma}^{G}=T_{\nu\mu\rho\sigma}^{G}\ ,\\
p^{\mu}T_{\mu\nu\rho\sigma}^{G}=0\ ,\\
g^{\mu\nu}T_{\mu\nu\rho\sigma}^{G}=0\ .
\end{array}
\right.
\end{equation}
As it is well-known, it is always possible to fulfill the first cited Ward
identity ($T_{\mu\nu\rho\sigma}^{G}=T_{\nu\mu\rho\sigma}^{G}$) by imposing
that the quantized energy-momentum tensor is symmetric \cite{BERTLMANN-1}.
Therefore, we need to investigate if the last two WIs can be satisfied also.
It is expected that both are broken by anomalous terms known as Einstein and
Weyl anomalies, respectively. In our investigation we will obtain, among other
things, a set of conditions to be fulfilled in order to satisfy these properties.

The main task of next sections is to check, after the explicit calculations of
$T_{\mu\nu\rho\sigma}^{G}$, if the obtained mathematical forms are, firstly,
in accordance with the RAGFs to, after this, verify if it is possible to
preserve the associated WIs. For the first task it is only required to be
careful in the operations, in order to obey the mathematics, while for the
second task it is expected that a set of additional conditions need to be
identified in order to be imposed in addition to the application of the
Feynman rules.

\section{The procedure for Handling Divergent Feynman Integrals
\label{sec_method}}

Most of QFT's predictions are made through perturbative methods. The
construction of the perturbative amplitudes, on the other hand, is
systematized by the well-known Feynman rules. Within this context we find,
invariably, a set of amplitudes at the loop level, corresponding to physical
processes, which are divergent quantities. This requires the adoption of an
adequate procedure in order to handle with this situation. Due to this reason,
in this section we present the procedure which we adopt to handle the
intrinsic mathematical problems of the perturbative series in QFTs. The
mathematical strategy adopted play a crucial role in our investigation.

In a first step, by applying the Feynman rules, we construct the desirable
perturbative amplitude, for one value of the loop momentum. Then, by a simple
power counting, it is stated the superficial degree of divergence. Therefore,
physical quantities, which are combinations of propagators and vertexes, may
be in an integrand of a divergent integral when the integration is taken over
the loop momentum, which, formally, corresponds to the implementation of the
last Feynman rule. The usual procedure is, at this point, to adopt a
regularization technique in order to make the integrals. This implies in to
modify the amplitudes as they come from the corresponding Feynman rules. After
all the operations are made, some limit is taken to, in principle, connect the
obtained results to the initial situation, removing the effects of the
mathematical modifications introduced. However, as it is well known, due to
the divergent character of the modified integrals, the integration and the
limit are not commuting operations such that the result is not unique and
(which is bad) is dependent on the intermediary sequence of steps followed.
Given this fact, some aspects of the perturbative calculations are prejudiced,
especially those were surface terms are involved since, in the regularized
expressions, they may assume prescription dependent values. In the dimensional
regularization (DR) prescription \cite{Hooft, Bollini, Ashmore}, as an
example, they are assumed as having vanishing values, allowing then shifts in
the integrating momentum. On the other hand, in prescriptions where the
regularization is made through distributions at a fixed space-time dimension,
the value for the referred surface terms are not zero and the amplitudes
became dependent in the particular routing adopted for the internal lines
momenta of the loop. Both methods produce very different implications in
qualitative and quantitative sense.

In order to avoid the previous described situation, it was developed a
procedure which can circumvent the modifications of the perturbative
amplitudes at the intermediary steps of the calculations such that all the
ingredients are present at the final form obtained, like the aspects related
to the surface terms involved. With this attitude a very rich analysis is
allowed since, as we have said, a correspondence with all specific
regularization technique is always possible.

The main idea is to assume that the linearity in the integration operation is
a valid property for Feynman integrals, in such a way it is possible to write
the expression for a perturbative amplitude in any mathematical form which is
mathematically identical to that usually adopted by the Feynman rules, before
implementing the last rule. Strictly speaking, there is an infinite number of
equivalent mathematical forms for the Feynman amplitudes. This freedom allow
us to choose the most convenient one for our purposes. We can assume, at this
point a criterion for the choice; the most simple mathematical expression
where\textit{ no physical parameters will be inside a divergent integral when
the last rule is implemented}. Our next task is, therefore, to rewrite the
propagators, where resides the dependence on the loop momentum, in a way which
allows us to achieve this goal.\ In principle, any identity which generates a
sequence of terms having a regressive power counting in the loop momentum can
be adopted. Probably the most simple one is the identity \cite{ORIMAR-TESE}%
\begin{align}
\frac{1}{D_{i}}  &  =\frac{1}{[(k+k_{i})^{2}-m_{i}^{2}]}\nonumber\\
&  =\sum_{j=0}^{N}\frac{\left(  -1\right)  ^{j}\left(  k_{i}^{2}+2k_{i}\cdot
k+\lambda^{2}-m_{i}^{2}\right)  ^{j}}{\left(  k^{2}-\lambda^{2}\right)
^{j+1}}\nonumber\\
&  +\frac{\left(  -1\right)  ^{N+1}\left(  k_{i}^{2}+2k_{i}\cdot k+\lambda
^{2}-m_{i}^{2}\right)  ^{N+1}}{\left(  k^{2}-\lambda^{2}\right)  ^{N+1}\left[
\left(  k+k_{i}\right)  ^{2}-m_{i}^{2}\right]  }\ , \label{prop_rep}%
\end{align}
where $N$ need to be taken as equal to or greater than the superficial degree
of divergence. On its turn, $\lambda$ is an arbitrary parameter having
dimension of mass and $k_{i}$ is an internal (arbitrary) momentum. In
practical terms, it is equivalent to say that the infinite forms allowed by
the value of $N$, in the above expression, are completely equivalent to
represent the required expression for a propagator in the application of
Feynman rules. It is enough that the linearity in the integration operation is
a valid mathematical property. In addition, it is required also that, in all
steps, the complete independence of the arbitrary parameter $\lambda$ is obtained.

The convenient use of the identity (\ref{prop_rep}) make possible to split up
any divergent Feynman integral into a sum of scalar (irreducible) divergent
integrals, surface terms and finite functions of the external momenta. The set
of divergent quantities is reduced to few objects which in our present
investigation is composed by two irreducible (scalar) ones%
\begin{align}
I_{\log}^{\left(  2\right)  }\left(  \lambda^{2}\right)   &  =\int\frac
{d^{2}k}{\left(  2\pi\right)  ^{2}}\frac{1}{\left(  k^{2}-\lambda^{2}\right)
}\ ,\\
I_{quad}^{\left(  2\right)  }\left(  \lambda^{2}\right)   &  =\int\frac
{d^{2}k}{\left(  2\pi\right)  ^{2}}\ln\left(  \frac{k^{2}-\lambda^{2}}{k^{2}%
}\right)  \ .
\end{align}
The masses within these objects (mass scales) can be changed freely through
identities that are called scale relations and stated by%
\begin{align}
\left[  I_{\log}^{\left(  2\right)  }\left(  \lambda^{2}\right)  \right]   &
=\left[  I_{\log}^{\left(  2\right)  }\left(  \lambda_{0}^{2}\right)  \right]
+\frac{i}{4\pi}\ln\left(  \frac{\lambda_{0}^{2}}{\lambda^{2}}\right)
\ ,\label{scale_I_log}\\
\left[  I_{quad}^{\left(  2\right)  }\left(  \lambda^{2}\right)  \right]   &
=\left[  I_{quad}^{\left(  2\right)  }\left(  \lambda_{0}^{2}\right)  \right]
+\left(  \lambda^{2}-\lambda_{0}^{2}\right)  \left[  I_{\log}^{\left(
2\right)  }\left(  \lambda_{0}^{2}\right)  \right] \nonumber\\
&  +\frac{i}{4\pi}\left[  \lambda^{2}-\lambda_{0}^{2}+\lambda^{2}\ln\left(
\frac{\lambda_{0}^{2}}{\lambda^{2}}\right)  \right]  \ . \label{scale_I_quad}%
\end{align}
On the other hand, we will find four quantities which can be recognized as
being surface terms%
\begin{align}
\Delta_{1;\mu\nu}^{\left(  2\right)  }  &  =\int\frac{d^{2}k}{\left(
2\pi\right)  ^{2}}\frac{\partial}{\partial k_{\mu}}\left[  k_{\nu}\left(
2-\ln\frac{k^{2}}{k^{2}-\lambda^{2}}\right)  \right] \nonumber\\
&  =\int\frac{d^{2}k}{\left(  2\pi\right)  ^{2}}\left\{  \frac{2k_{\mu}k_{\nu
}}{\left(  k^{2}-\lambda^{2}\right)  }-g_{\mu\nu}\ln\left(  \frac{k^{2}}%
{k^{2}-\lambda^{2}}\right)  \right\}  \ , \label{Delta1}%
\end{align}%
\begin{align}
\Delta_{2;\mu\nu}^{\left(  2\right)  }  &  =\int\frac{d^{2}k}{\left(
2\pi\right)  ^{2}}\frac{\partial}{\partial k_{\mu}}\left(  -\frac{k_{\nu}%
}{\left(  k^{2}-\lambda^{2}\right)  }\right) \nonumber\\
&  =\int\frac{d^{2}k}{\left(  2\pi\right)  ^{2}}\left\{  \frac{2k_{\mu}k_{\xi
}}{\left(  k^{2}-\lambda^{2}\right)  ^{2}}-\frac{g_{\mu\xi}}{\left(
k^{2}-\lambda^{2}\right)  }\right\}  \ , \label{Delta2}%
\end{align}%
\begin{align}
\square_{2;\mu\nu\alpha\beta}^{\left(  2\right)  }  &  =\int\frac{d^{2}%
k}{\left(  2\pi\right)  ^{2}}\frac{\partial}{\partial k_{\mu}}\left(
-\frac{2k_{\nu}k_{\alpha}k_{\beta}}{\left(  k^{2}-\lambda^{2}\right)  }\right)
\nonumber\\
&  =\int\frac{d^{2}k}{\left(  2\pi\right)  ^{2}}\left\{  \frac{4k_{\mu}k_{\nu
}k_{\alpha}k_{\beta}}{\left(  k^{2}-\lambda^{2}\right)  ^{2}}-g_{\mu\nu}%
\frac{2k_{\alpha}k_{\beta}}{\left(  k^{2}-\lambda^{2}\right)  }\right.
\nonumber\\
&  \left.  -g_{\mu\alpha}\frac{2k_{\nu}k_{\beta}}{\left(  k^{2}-\lambda
^{2}\right)  }-g_{\mu\beta}\frac{2k_{\nu}k_{\alpha}}{\left(  k^{2}-\lambda
^{2}\right)  }\right\}  \ , \label{Box1}%
\end{align}%
\begin{align}
\square_{3;\mu\nu\alpha\beta}^{\left(  2\right)  }  &  =\int\frac{d^{2}%
k}{\left(  2\pi\right)  ^{2}}\frac{\partial}{\partial k_{\mu}}\left(
-\frac{2k_{\nu}k_{\alpha}k_{\beta}}{\left(  k^{2}-\lambda^{2}\right)  ^{2}%
}\right) \nonumber\\
&  =\int\left\{  \frac{d^{2}k}{\left(  2\pi\right)  ^{2}}\frac{8k_{\mu}k_{\nu
}k_{\alpha}k_{\beta}}{\left(  k^{2}-\lambda^{2}\right)  ^{3}}-g_{\mu\nu}%
\frac{2k_{\alpha}k_{\beta}}{\left(  k^{2}-\lambda^{2}\right)  ^{2}}\right.
\nonumber\\
&  \left.  -g_{\mu\alpha}\frac{2k_{\nu}k_{\beta}}{\left(  k^{2}-\lambda
^{2}\right)  ^{2}}-g_{\mu\beta}\frac{2k_{\nu}k_{\alpha}}{\left(  k^{2}%
-\lambda^{2}\right)  ^{2}}\right\}  \ , \label{Box2}%
\end{align}%
\begin{align}
\Sigma_{4;\mu\nu\alpha\beta\xi\chi}^{\left(  2\right)  }  &  =\int\frac
{d^{2}k}{\left(  2\pi\right)  ^{2}}\frac{\partial}{\partial k_{\mu}}\left(
-\frac{8k_{\nu}k_{\alpha}k_{\beta}k_{\xi}k_{\chi}}{\left(  k^{2}-\lambda
^{2}\right)  ^{3}}\right) \nonumber\\
&  =\int\frac{d^{2}k}{\left(  2\pi\right)  ^{2}}\left\{  \frac{48k_{\mu}%
k_{\nu}k_{\alpha}k_{\beta}k_{\xi}k_{\chi}}{\left(  k^{2}-\lambda^{2}\right)
^{4}}-g_{\mu\nu}\frac{8k_{\alpha}k_{\beta}k_{\xi}k_{\chi}}{\left(
k^{2}-\lambda^{2}\right)  ^{3}}\right. \nonumber\\
&  -g_{\mu\alpha}\frac{8k_{\nu}k_{\beta}k_{\xi}k_{\chi}}{\left(  k^{2}%
-\lambda^{2}\right)  ^{3}}-g_{\mu\beta}\frac{8k_{\nu}k_{\alpha}k_{\xi}k_{\chi
}}{\left(  k^{2}-\lambda^{2}\right)  ^{3}}\nonumber\\
&  \left.  -g_{\mu\xi}\frac{8k_{\nu}k_{\alpha}k_{\beta}k_{\chi}}{\left(
k^{2}-\lambda^{2}\right)  ^{3}}-g_{\mu\chi}\frac{8k_{\nu}k_{\alpha}k_{\beta
}k_{\xi}}{\left(  k^{2}-\lambda^{2}\right)  ^{3}}\right\}  \ . \label{Sigma}%
\end{align}
The convenience of this systematization will be clear in future discussions.
In turn, the finite integrals arising can be integrated out without
restrictions and the results written in terms of a set of finite functions
defined through integral representations given in terms of Feynman parameters
\cite{ORIMAR-SIST-1}. In the present work such structures are defined by
\begin{align}
\xi_{k}^{\left(  -1\right)  }\left(  p^{2},m^{2}\right)   &  =\int_{0}%
^{1}dx\frac{x^{k}}{Q\left(  p^{2},m^{2};x\right)  }\ ,\\
\xi_{k}^{\left(  0\right)  }\left(  p^{2},m^{2};\lambda^{2}\right)   &
=\int_{0}^{1}dx\ x^{k}\ln\left[  \frac{Q\left(  p^{2},m^{2};x\right)
}{-\lambda^{2}}\right]  \ ,
\end{align}
with $k=0,1,2,...$ and the polynomial $Q$ given by $Q\left(  p^{2}%
,m^{2};x\right)  =p^{2}x(1-x)-m^{2}$. There are, obviously, relations between
the functions corresponding to two different values of the index $k$ ,
allowing us to reduce them to the $\xi_{0}^{\left(  -1\right)  }$ or $\xi
_{0}^{\left(  0\right)  }$, which is particularly useful in RAGFs or WIs verification.

Observe that within a traditional regularization prescription the divergent
objects would have a value attributed to them. For example, in DR all the
above surface terms are taken as being zero and the $I_{\log}\left(
\lambda^{2}\right)  $ and $I_{quad}\left(  \lambda^{2}\right)  $ objects
manifest themselves as poles, for specific values of the space-time dimension,
in the amplitudes. In fact, one can always formulate a one-to-one map among
our results and those produced by regularizations prescriptions, as will
become clear in what follows. On the other hand, in our prescription, they
remain untouched and are present in the final results, where their possible
values could be considered and tested for consistency requirements. These
aspects represents the heart of the analysis and conclusions made in this job.

\section{Explicitly evaluation of the subamplitudes and the verification of
their RAGFs \label{sec_exp_eval}}

In order to explicitly calculate the gravitational amplitude $T_{\mu\nu
\rho\sigma}^{G}$, we first calculate its subamplitudes, through the strategy
described above, after that we check whether the results obtained are
consistent with the corresponding RAGFs and, if it is the case, verify if it
is possible to satisfy the WIs.

We start with the subamplitudes composing the vector sector given by%
\begin{align}
T_{\mu\nu\rho\sigma}^{\left(  V\right)  }  &  =4\left[  T_{\nu\sigma;\mu\rho
}^{VV}\right]  +2p_{\sigma}\left[  T_{\nu;\mu\rho}^{VV}\right]  +2p_{\nu
}\left[  T_{\sigma;\mu\rho}^{VV}\right]  +p_{\nu}p_{\sigma}\left[  T_{\mu\rho
}^{VV}\right] \nonumber\\
&  +\left(  \mu\longleftrightarrow\nu\right)  +\left(  \sigma
\longleftrightarrow\rho\right)  \ .
\end{align}
These three kind of subamplitudes are defined by taking $\Gamma_{i}%
=\gamma_{\mu}$ and $\Gamma_{j}=\gamma_{\nu}$ in definitions (\ref{T2}),
(\ref{T3}), and (\ref{T4}).

\subsection{$T_{\sigma\rho}^{VV}$ amplitude}

Let us consider first the $T_{\mu\nu}^{VV}$ (see Eq. (\ref{T2})). This
amplitude belongs to the spectrum of amplitudes arising in renormalizable
theories like $QED_{2}$, where the gauge invariance plays a crucial role. In
such a context it represents the one-loop polarization tensor. Therefore, its
identification as a substructure of the $T_{\mu\nu\rho\sigma}^{G}$ amplitude
may shine some light in our investigation about gravitational anomalies. We
will consider, due to this, some details in the procedures.

We know that it is expected to identify the relations (\ref{VV_RAGF_1}) and
(\ref{VV_RAGF_2}) as properties of its explicit form. Thus, it is relevant to
know the corresponding expression for the one-point vector function in
advance. Beside that, it is a good opportunity to exemplify the use of the
procedure in a simple algebraic scenario. First, after taking the Dirac
traces, we obtain the expression for one value of the loop momentum%
\begin{equation}
t_{\mu}^{V}\left(  k_{1}\right)  =2\left[  \frac{k^{\alpha}}{D_{1}}%
+k_{1}^{\alpha}\frac{1}{D_{1}}\right]  \ .
\end{equation}
For the first term we adopt for the propagator the representation%
\begin{align}
\frac{k_{\mu}}{D_{1}}  &  =\frac{k_{\mu}}{\left(  k^{2}-\lambda^{2}\right)
}-\frac{\left(  k_{1}^{2}+2k\cdot k_{1}+\lambda^{2}-m^{2}\right)  k_{\mu}%
}{\left(  k^{2}-\lambda^{2}\right)  ^{2}}\nonumber\\
&  +\frac{\left(  k_{1}^{2}+2k\cdot k_{1}+\lambda^{2}-m^{2}\right)  ^{2}%
k_{\mu}}{\left(  k^{2}-\lambda^{2}\right)  ^{2}\left[  \left(  k+k_{1}\right)
^{2}-m^{2}\right]  }\ ,
\end{align}
which corresponds to adopt $N=1$ in (\ref{prop_rep}). For the second term the
same representation can be adopted. However, in order to avoid unnecessary
algebraic efforts, one can take $N=0$,
\begin{equation}
\frac{1}{D_{1}}=\frac{1}{\left(  k^{2}-\lambda^{2}\right)  }-\frac{\left(
k_{1}^{2}+2k\cdot k_{1}+\lambda^{2}-m^{2}\right)  }{\left(  k^{2}-\lambda
^{2}\right)  \left[  \left(  k+k_{1}\right)  ^{2}-m^{2}\right]  }.
\end{equation}
Thus
\begin{align}
\left[  \frac{k_{\alpha}}{D_{1}}+{k}_{1\alpha}\frac{1}{D_{1}}\right]
_{not\text{ }odd}  &  =-{k}_{1}^{\beta}\left\{  \frac{2k_{\alpha}k_{\beta}%
}{\left(  k^{2}-\lambda^{2}\right)  ^{2}}-\frac{g_{\alpha\beta}}{\left(
k^{2}-\lambda^{2}\right)  }\right\} \nonumber\\
&  +\frac{\left(  k_{1}^{2}+2k\cdot k_{1}+\lambda^{2}-m^{2}\right)  ^{2}%
}{\left(  k^{2}-\lambda^{2}\right)  ^{2}\left[  \left(  k+k_{1}\right)
^{2}-m^{2}\right]  }k_{\alpha}\nonumber\\
&  -\frac{\left(  k_{1}^{2}+2k\cdot k_{1}+\lambda^{2}-m^{2}\right)  }{\left(
k^{2}-\lambda^{2}\right)  \left[  \left(  k+k_{1}\right)  ^{2}-m^{2}\right]
}{k}_{1\alpha}\ ,
\end{align}
where an odd term was omitted since, after the integration, it will vanishes.
Note that the dependence in the arbitrary internal momentum is located only in
finite integrals. The divergent terms will not contain physical quantities
since $\lambda$ is an arbitrary parameter. After some reorganization, we can
take the integration on both sides
\begin{align}
&  \int\frac{d^{2}k}{\left(  2\pi\right)  ^{2}}\left[  t_{\mu}^{V}\left(
k_{1}\right)  +{k}_{1}^{\beta}\left\{  \frac{2k_{\alpha}k_{\beta}}{\left(
k^{2}-\lambda^{2}\right)  ^{2}}-\frac{g_{\alpha\beta}}{\left(  k^{2}%
-\lambda^{2}\right)  }\right\}  \right] \nonumber\\
&  =\int\frac{d^{2}k}{\left(  2\pi\right)  ^{2}}\frac{k_{\alpha}\left(
k_{1}^{2}+2k\cdot k_{1}+\lambda^{2}-m^{2}\right)  ^{2}}{\left(  k^{2}%
-\lambda^{2}\right)  ^{2}\left[  \left(  k+k_{1}\right)  ^{2}-m^{2}\right]
}\nonumber\\
&  -{k}_{1\alpha}\int\frac{d^{2}k}{\left(  2\pi\right)  ^{2}}\frac{\left(
k_{1}^{2}+2k\cdot k_{1}+\lambda^{2}-m^{2}\right)  }{\left(  k^{2}-\lambda
^{2}\right)  \left[  \left(  k+k_{1}\right)  ^{2}-m^{2}\right]  }\ .
\end{align}
This expression is only consequence of the Feynman rules. On the right hand
side, there are only finite integrals. They can be solved without any kind of
concern since any reasonable regularization must give the same result for a
finite integral. The integration reveals the identically zero value. A formal
relation can be written by identifying a surface term like the one defined in
(\ref{Delta2}),
\begin{equation}
T_{\mu}^{V}\left(  k_{1}\right)  =-2k_{1}^{\alpha}\left[  \Delta_{2;\mu\alpha
}^{\left(  2\right)  }\left(  \lambda^{2}\right)  \right]  \ . \label{res_V}%
\end{equation}
The result is proportional to the arbitrary momentum $k_{1}$ and to the
surface term $\Delta_{2;\mu\alpha}^{\left(  2\right)  }$, whose argument is
also an arbitrary quantity. The mathematical object $\Delta_{2;\mu\alpha
}^{\left(  2\right)  }$ is prescription dependent in the sense that, in order
to attribute a definite value to it, some particular mathematical procedure is
required.\ It would be desirable that this perturbative amplitude gives a null
result, as we can see below, but the Feynman rules do not imply that.

The same procedure can be used to evaluate the $T_{\sigma\rho}^{VV}$
amplitude. The result can be written as%
\begin{equation}
T_{\sigma\rho}^{VV}=2\left[  \Delta_{2;\sigma\rho}^{\left(  2\right)
}\right]  +\frac{i}{\pi}\left(  p_{\sigma}p_{\rho}-g_{\sigma\rho}p^{2}\right)
\left[  \xi_{2}^{\left(  -1\right)  }\left(  p^{2};m^{2}\right)  -\xi
_{1}^{\left(  -1\right)  }\left(  p^{2};m^{2}\right)  \right]  \ ,
\label{res_VV2}%
\end{equation}
where $p = k_2 - k_1$. In the above result we can see clearly the aforementioned organization through
finite and (a priori) divergent objects. For our purposes, it is important to
known if the above result is in accordance to the expected RAGFs (Eqs.
(\ref{VV_RAGF_1}) and (\ref{VV_RAGF_2})). The contraction of (\ref{res_VV2})
with $p^{\sigma}$ gives%
\begin{equation}
p^{\sigma}T_{\sigma\rho}^{VV}=2\left(  {k_{2}}-k_{1}\right)  ^{\sigma}\left[
\Delta_{2;\sigma\rho}^{\left(  2\right)  }\right]  \ ,
\end{equation}
which, given (\ref{res_V}), can be recognized as being the RAGFs
(\ref{VV_RAGF_1}). So, this (vector) RAGFs is automatically fulfilled by
(\ref{res_VV2}). The same conclusion is also valid for $p^{\rho}$ contraction.

This result is a good opportunity to illustrate our preceding comments about
RAGFs and WIs. Since the $T_{\sigma\rho}^{VV}$ amplitude is proportional to
the polarization tensor of $QED_{2}$, gauge invariance implies that it must
have two conserved vector currents, i.e., $p^{\sigma}T_{\sigma\rho}%
^{VV}=p^{\rho}T_{\sigma\rho}^{VV}=0$. It is easy to see that such requirements
is not automatically satisfied by expression (\ref{res_VV2}). The violating
term is given by the (undefined) object $\Delta_{2;\sigma\rho}^{\left(
2\right)  }$, according to the expectations. The Feynman rules ended their
job. The RAGFs\ are satisfied as it is required but the WIs satisfaction will
depend on an additional ingredient. So, without additional assumptions, the
unique way to obtain a polarization tensor $T_{\sigma\rho}^{VV}$ satisfying
its WIs is in the absence of the object $\Delta_{2;\sigma\rho}^{\left(
2\right)  }$. As a surface term, this is exactly what would happens if we have
used the DR prescription. This is, in fact, a necessary requirement for all
regularizations which intend to be gauge preserving \cite{ORIMAR-PRD1}.

\subsection{$T_{\mu;\sigma\rho}^{VV}$ amplitude}

The next subamplitude to consider is $T_{\mu;\sigma\rho}^{VV}$ (see Eq.
(\ref{T3})). The result can be put in the form%
\begin{align}
T_{\mu;\sigma\rho}^{VV}  &  =-P^{\alpha}\left[  \square_{3;\alpha\mu\sigma
\rho}^{\left(  2\right)  }\right]  +P_{\rho}\left[  \Delta_{2;\sigma\mu
}^{\left(  2\right)  }\right]  +P_{\sigma}\left[  \Delta_{2;\rho\mu}^{\left(
2\right)  }\right] \nonumber\\
&  +P^{\alpha}\left\{  g_{\sigma\rho}\left[  \Delta_{2;\mu\alpha}^{\left(
2\right)  }\right]  -g_{\mu\rho}\left[  \Delta_{2;\alpha\sigma}^{\left(
2\right)  }\right]  -g_{\mu\sigma}\left[  \Delta_{2;\rho\alpha}^{\left(
2\right)  }\right]  \right\} \nonumber\\
&  -\frac{p_{\mu}}{2}\left[  T_{\sigma\rho}^{VV}\right]  \ , \label{res_VV3}%
\end{align}
where $P = k_2 + k_1$. The proposed systematization is, again, clear from the above expression. In
the above equation we also see the polarization tensor $T_{\sigma\rho}^{VV}$
as being a substructure of $T_{\mu;\sigma\rho}^{VV}$. The crucial question is:
does the above expression fulfill its expected RAGFs?

The contraction of the above result with $p^{\sigma}$ reveals that the RAGFs
(\ref{RAGF_VV_3_index_p}) is satisfied. The same happens for the $p^{\mu}$
contraction. These calculations can easily be done by observing the results
for the one-point amplitude $T_{\mu;\nu}^{V}\left(  k_{i}\right)  $,\ given in
Eq. (\ref{One_point_V_2_index}). On the other hand, the metric contraction
gives%
\begin{align}
g^{\mu\sigma}\left[  T_{\mu;\sigma\rho}^{VV}\right]   &  =\left[  T_{\rho}%
^{V}\left(  k_{2}\right)  \right]  +m\left[  T_{\rho}^{SV}\right] \nonumber\\
&  -\left(  k_{2}+k_{1}\right)  ^{\sigma}\left\{  \left[  g^{\mu\nu}%
\square_{3;\mu\nu\sigma\rho}^{\left(  2\right)  }\right]  -g_{\sigma\rho
}\left[  g^{\mu\nu}\Delta_{2;\mu\nu}^{\left(  2\right)  }\right]  \right\}
\ ,
\end{align}
where $T_{\rho}^{SV}$ is given by Eq.\ (\ref{two_point_SV_1_index}). The last
term of the above equation shows that expression (\ref{res_VV3}) for
$T_{\mu;\sigma\rho}^{VV}$ does not (automatically) satisfy the expect RAGFs
(\ref{RAGF_VV_3_index_g}). The spurious terms are composed by two surface
terms. In order to not break this RAGFs we must have%
\begin{equation}
g^{\mu\nu}\square_{3;\mu\nu\sigma\rho}^{\left(  2\right)  }=g_{\sigma\rho
}\left[  g^{\mu\nu}\Delta_{2;\mu\nu}^{\left(  2\right)  }\right]  \ .
\label{req_1}%
\end{equation}
Let us consider this in an explicit way. First we note that the integrand of
$g^{\mu\nu}\Delta_{2;\mu\nu}^{\left(  2\right)  }$ satisfy a trivial algebraic
identity\emph{ }%
\begin{equation}
\frac{2k^{2}}{\left(  k^{2}-m^{2}\right)  ^{2}}-\frac{2}{\left(  k^{2}%
-m^{2}\right)  }=\frac{2m^{2}}{\left(  k^{2}-m^{2}\right)  ^{2}},
\end{equation}
which means that the integral is finite, as well as the linearity of the
integration operation is assumed. Given that one obtain%
\begin{equation}
g^{\mu\nu}\Delta_{2;\mu\nu}^{\left(  2\right)  }=-\frac{i}{2\pi}.
\label{trace_Delta_2}%
\end{equation}
In a similar way, the integrand of the quantity $g^{\mu\nu}\left[
\square_{3;\mu\nu\sigma\rho}\right]  $ can be rewritten as%
\begin{equation}
\frac{8k^{2}k_{\sigma}k_{\rho}}{\left(  k^{2}-m^{2}\right)  ^{3}}%
-\frac{8k_{\sigma}k_{\rho}}{\left(  k^{2}-m^{2}\right)  ^{2}}=\frac
{8m^{2}k_{\sigma}k_{\rho}}{\left(  k^{2}-m^{2}\right)  ^{3}},
\end{equation}
such that the corresponding integral will be finite. Performing the
integration we get%
\begin{equation}
g^{\mu\nu}\left[  \square_{3;\mu\nu\sigma\rho}\right]  =-\frac{i}{2\pi
}g_{\sigma\rho}\ .
\end{equation}
Given both results we obtain (\ref{req_1}). This means that our procedure is
consistent with the linearity in the integration operation. It is interesting
to note that we have evaluated surface terms and the results obtained are
nonzero. Here one can note that the condition (\ref{req_1}) would also be
satisfied by assuming $\square_{3;\mu\nu\sigma\rho}^{\left(  2\right)
}=\Delta_{2;\mu\nu}^{\left(  2\right)  }=0$. If we had applied the DR to
perform these calculations, such requirements would be fulfilled automatically
since in the DR prescription surface terms are assumed to vanish.

\subsection{$T_{{\mu\nu;\sigma\rho}}^{VV}$ amplitude}

The last subamplitude composing the vector sector is $T_{{\mu\nu;\sigma\rho}%
}^{VV}$ (see Eq. (\ref{T4})). After a long and tedious but a straightforward
calculation, we obtain%
\begin{align}
T_{{\mu\nu;\sigma\rho}}^{VV}  &  =S_{{\mu\nu;\sigma\rho}}^{VV}+\left(
g_{\rho\mu}g_{\sigma\nu}+g_{\rho\nu}g_{\sigma\mu}\right)  \left[
I_{quad}^{\left(  2\right)  }\left(  m^{2}\right)  \right] \nonumber\\
&  +\left\{  \frac{1}{3}\left(  g_{\mu\nu}g_{\rho\sigma}p^{2}-g_{\mu\nu
}p_{\rho}p_{\sigma}-g_{\sigma\rho}p_{\mu}p_{\nu}\right)  \right. \nonumber\\
&  \text{ \ }\ \ \ \ -\frac{1}{6}\left(  g_{\mu\rho}g_{\nu\sigma}p^{2}%
-g_{\mu\rho}p_{\nu}p_{\sigma}-g_{\nu\sigma}p_{\mu}p_{\rho}\right) \nonumber\\
&  \text{ \ \ \ \ }\left.  -\frac{1}{6}\left(  g_{\nu\rho}g_{\mu\sigma}%
p^{2}-g_{\nu\rho}p_{\mu}p_{\sigma}-g_{\mu\sigma}p_{\nu}p_{\rho}\right)
\right\}  \left[  I_{\log}^{\left(  2\right)  }\left(  m^{2}\right)  \right]
\nonumber\\
&  +\frac{i}{2\pi}\left\{  \frac{p_{\mu}p_{\nu}p_{{\sigma}}p_{\rho}}{p^{2}%
}+\frac{p^{2}}{2}\left(  g_{\nu{\rho}}g_{\mu{\sigma}}+g_{\nu\sigma}g_{\mu
{\rho}}\right)  \right. \nonumber\\
&  \text{ \ \ \ \ \ \ \ \ \ \ \ }\left.  -\frac{p_{\nu}}{2}\left(
g_{\mu\sigma}p_{\rho}+g_{\mu\rho}p_{\sigma}\right)  -\frac{p_{\mu}}{2}\left(
g_{\nu{\rho}}p_{{\sigma}}+g_{\nu{\sigma}}p_{{\rho}}\right)  \right\}  \left[
2\xi_{2}^{\left(  0\right)  }-\xi_{1}^{\left(  0\right)  }\right] \nonumber\\
&  +\frac{i}{2\pi}\frac{1}{p^{2}}\left(  p_{\sigma}p_{\rho}-p^{2}%
g_{{\sigma\rho}}\right)  \left(  p_{\mu}p_{\nu}-p^{2}g_{\mu\nu}\right)
\left[  \xi_{2}^{\left(  0\right)  }-\xi_{1}^{\left(  0\right)  }\right]
+\frac{1}{4}p_{\rho}p_{\sigma}\left[  T_{\mu\nu}^{VV}\right]  \ ,
\label{res_VV4}%
\end{align}
where $S_{{\mu\nu;\sigma\rho}}^{VV}$ represents a (ambiguous) combination of
surface terms and is given explicitly in appendix (\ref{Ap_2P_Func}). Note
that in the expression above the divergent objects appear as functions of the
physical mass ($m$) rather than an arbitrary scale mass ($\lambda$). From now
on, we will adopt this simplified notation, because the mass scale, chosen for
the divergent objects, will not play an important role in the discussions
presented in this work. In addition, if needed for some reason, the mass scale
can be changed freely using the scale relations shown in the Eqs.
(\ref{scale_I_log}) and (\ref{scale_I_quad}). The contraction of the above
result with $p^{\sigma}$ or $p^{\mu}$ shows that both expected RAGFs (see Eqs.
(\ref{RAGF_VV_4_index_p1}) and (\ref{RAGF_VV_4_index_p2})), are satisfied
automatically while the contraction with the metric gives%
\begin{align}
g^{\mu\sigma}\left[  T_{{\mu\nu;\sigma\rho}}^{VV}\right]   &  =\left[
T_{\nu;\rho}^{V}\left(  k_{2}\right)  \right]  +m\left[  T_{\nu;\rho}%
^{SV}\right] \nonumber\\
&  +\frac{1}{3}\left(  k_{2}^{\xi}k_{2}^{\chi}+k_{1}^{\xi}k_{2}^{\chi}%
+k_{1}^{\xi}k_{1}^{\chi}\right)  \left[  g^{\mu\sigma}\Sigma_{4;\mu\sigma
\nu\rho\xi\chi}^{\left(  2\right)  }\right] \nonumber\\
&  -\frac{1}{2}\left(  k_{1}^{2}+k_{2}^{2}\right)  \left[  g^{\mu\sigma
}\square_{3;\mu\sigma\nu\rho}^{\left(  2\right)  }\right] \nonumber\\
&  -\frac{1}{2}\left(  {k_{2}}+{k_{1}}\right)  _{{\rho}}\left(  k_{2}%
+k_{1}\right)  ^{\xi}\left[  g^{\mu\sigma}\square_{3;\mu\sigma\nu\xi}^{\left(
2\right)  }\right] \nonumber\\
&  -\left(  k_{2}+k_{1}\right)  ^{\xi}k_{1{\nu}}\left[  g^{\mu\sigma}%
\square_{3;\mu\sigma\rho\xi}^{\left(  2\right)  }\right]  +\left(  {k_{2}%
}+{k_{1}}\right)  _{{\rho}}k_{1{\nu}}\left[  g^{\mu\sigma}\Delta_{2;\mu\sigma
}^{\left(  2\right)  }\right] \nonumber\\
&  +\frac{i}{4\pi}\frac{1}{3}\left(  p_{{\rho}}p_{\nu}-g_{\nu{\rho}}%
p^{2}\right)  \ ,
\end{align}
where we have used the results (\ref{One_point_V_2_index}) and
(\ref{two_point_SV_2_index}). Now there are six potentially breaking terms for
the RAGFs (\ref{RAGF_VV_4_index_g}). In order to fulfill this RAGFs we have to
get fulfilled both the condition (\ref{req_1}) and also
\begin{equation}
g^{\mu\sigma}\Sigma_{4;\mu\sigma\nu\rho\xi\chi}^{\left(  2\right)  }=-\frac
{i}{2\pi}\left(  g_{\nu\rho}g_{\xi\chi}+g_{\nu\xi}g_{\rho\chi}+g_{\nu\chi
}g_{\rho\xi}\right)  \ .
\end{equation}
It is simple to see that this condition is satisfied by using the same
sequence of steps used to obtain (\ref{req_1}), i.e., by assuming the validity
of the linearity in the integration operation. This result means that the
RAGFs is preserved by the operations made.

\subsection{$T_{\sigma\rho}^{AV}$ amplitude}

In the axial-vector sector we can make a similar investigation of the
pertinent set of subamplitudes, which are defined by taking $\Gamma_{i}%
=\gamma_{\sigma}\gamma_{3}$ and $\Gamma_{j}=\gamma_{\rho}$ in (\ref{T2}),
(\ref{T3}), and (\ref{T4}).

Let us take the simplest amplitude of the set, namely $T_{\sigma\rho}^{AV}$.
This subamplitude is a very interesting one for our present investigation
since it is the well-known \textit{anomalous} amplitude belonging to the
chiral $QED_{2}$ \cite{BERTLMANN-BOOK,ADAM,JOHNSON,ABDALLA,FUJIKAWA}. Its
evaluation can be made trivial if one note the relation
\begin{equation}
T_{\sigma\rho}^{AV}=-\varepsilon_{\sigma\alpha}g^{\alpha\beta}\left[
T_{\beta\rho}^{VV}\right]  \ , \label{AV_VV}%
\end{equation}
such that, by using (\ref{res_VV2}), we get%
\begin{equation}
T_{\sigma\rho}^{AV}=-2\varepsilon_{\sigma\xi}\left[  \Delta_{2;\xi\rho
}^{\left(  2\right)  }\right]  -\frac{i}{\pi}\varepsilon_{\sigma\xi}\left(
p_{\xi}p_{\rho}-g_{\xi\rho}p^{2}\right)  \left[  \xi_{2}^{\left(  -1\right)
}\left(  p^{2};m^{2}\right)  -\xi_{1}^{\left(  -1\right)  }\left(  p^{2}%
;m^{2}\right)  \right]  \ . \label{res_AV2}%
\end{equation}
There are two RAGFs expected to be satisfied by the above expression, which
were stated in (\ref{RAGF_AV_2_index_p1}) and (\ref{RAGF_AV_2_index_p2}). The
first one refers to the contraction with $p^{\rho}$. This contraction gives,
immediately, the expected difference $T_{{\nu}}^{A}\left(  k_{1}\right)
-T_{{\nu}}^{A}\left(  k_{2}\right)  $ (see Eq. (\ref{One_point_A_1_index})).

On the other hand, the contraction of (\ref{res_AV2}) with the axial index
($p^{\sigma}$) reveals%
\begin{align}
p^{\sigma}T_{\sigma\rho}^{AV}  &  =-2\varepsilon_{\sigma\xi}\left(
k_{2}-k_{1}\right)  ^{\sigma}g^{\xi\chi}\left[  \Delta_{2;\chi\rho}^{\left(
2\right)  }\right] \nonumber\\
&  -\frac{i}{\pi}\varepsilon_{\sigma\rho}p^{\sigma}\left[  1+m^{2}\xi
_{0}^{\left(  -1\right)  }\left(  p^{2},m^{2}\right)  \right]  \ .
\end{align}
Now, there is subtlety in order to identify the expected difference of two
axial one-point functions in the right hand side of the equation above. This
is a very important aspect of our investigation. For this, it is first
necessary to change the position of the Lorentz indexes in the first term
above. Through the Schouten identity%
\begin{equation}
\varepsilon_{\sigma\xi}g^{\xi\chi}\left[  \Delta_{2;\chi\rho}^{\left(
2\right)  }\right]  =\varepsilon_{\rho\xi}g^{\xi\chi}\left[  \Delta
_{2;\chi\sigma}^{\left(  2\right)  }\right]  -\varepsilon_{\sigma\rho}\left[
g^{\xi\chi}\Delta_{2;\xi\chi}^{\left(  2\right)  }\right]  \ ,
\label{Schouten}%
\end{equation}
such a change of index can be achieved. Given the $PV$ amplitude (see Eq.
(\ref{two_point_PV_1_index})), we can write%
\begin{align}
p^{\sigma}T_{\sigma\rho}^{AV}  &  =\left[  T_{\rho}^{A}\left(  k_{1}\right)
\right]  -\left[  T_{\rho}^{A}\left(  k_{2}\right)  \right]  +2m\left[
T_{\rho}^{PV}\right] \nonumber\\
&  +2\varepsilon_{\rho\xi}p^{\xi}\left\{  \frac{i}{2\pi}+\left[
g^{\alpha\beta}\Delta_{2;\alpha\beta}^{\left(  2\right)  }\right]  \right\}
\ .
\end{align}
From the expression above we see that the RAGFs (\ref{RAGF_AV_2_index_p2}) is
preserved, if and only if,%
\begin{equation}
g^{\xi\chi}\Delta_{2;\xi\chi}^{\left(  2\right)  }=-\frac{i}{2\pi}\ ,
\end{equation}
which is the same result founded in (\ref{trace_Delta_2}). Again, the
operations made are in accordance to the linearity in the integration
operation. However, it must be noted that through the Schouten identity we
have constructed two representations for the $AV$ amplitude, such that, after
the immediately above result, \ allow us to identify%
\begin{equation}
p^{\rho}\left[  T_{\sigma\rho}^{AV}\right]  _{1}=T_{{\nu}}^{A}\left(
k_{1}\right)  -T_{{\nu}}^{A}\left(  k_{2}\right)  \ ,
\end{equation}%
\begin{equation}
p^{\sigma}\left[  T_{\sigma\rho}^{AV}\right]  _{2}=\left[  T_{\rho}^{A}\left(
k_{1}\right)  \right]  -\left[  T_{\rho}^{A}\left(  k_{2}\right)  \right]
+2m\left[  T_{\rho}^{PV}\right]
\end{equation}
Both expressions $\left[  T_{\sigma\rho}^{AV}\right]  _{1}$ and $\left[
T_{\sigma\rho}^{AV}\right]  _{2}$ are identical from the mathematical point of
view, as long as the Schouten identity is valid. Note that the referred
identity relates a tensor with its trace.

It is possible to note an interesting aspect in the above results.\emph{ }In
the amplitudes of the vector sector we evaluated the finite quantities
$g^{\xi\chi}\Delta_{2;\xi\chi}^{\left(  2\right)  }$, $g^{\mu\nu}%
\square_{3;\mu\nu\rho\xi}^{\left(  2\right)  }$, and $g^{\mu\sigma}%
\Sigma_{4;\mu\sigma\nu\rho\xi\chi}^{\left(  2\right)  }$.\emph{ }In those
cases, the nonzero value found for these quantities put the results in
accordance with the requirements but the null value fulfill the conditions as
well. In the present case, the value $g^{\xi\chi}\Delta_{2;\xi\chi}^{\left(
2\right)  }=0$ breaks the linearity in the integration operation since the
axial index contraction gives%
\begin{equation}
p^{\sigma}T_{\sigma\rho}^{AV}=T_{{\rho}}^{A}\left(  k_{1}\right)  -T_{{\rho}%
}^{A}\left(  k_{2}\right)  +2m\left[  T_{\rho}^{PV}\right]  +\frac{i}{\pi
}\varepsilon_{\rho\xi}p^{\xi}\ .
\end{equation}
On the other hand, in order to fulfill the WIs
\begin{equation}
\left\{
\begin{array}
[c]{c}%
p^{\rho}T_{\sigma\rho}^{AV}=0\ ,\\
p^{\sigma}T_{\sigma\rho}^{AV}=2m\left[  T_{\rho}^{PV}\right]  \ ,
\end{array}
\right.
\end{equation}
a necessary condition is
\begin{equation}
T_{{\rho}}^{A}\left(  k_{1}\right)  -T_{{\rho}}^{A}\left(  k_{2}\right)  =0\ ,
\end{equation}
which is guaranteed only by some prescription that attributes a null value for
the object $\Delta_{2;\xi\chi}^{\left(  2\right)  }$, in the same way as it
was required in the case of the WIs for the polarization tensor ($T_{\sigma
\rho}^{VV}$). So, it seems that it is apparently possible to preserve the
linearity in the integration operation and both WIs, simultaneously, with the
conditions%
\begin{equation}
\left\{
\begin{array}
[c]{c}%
\Delta_{2;\xi\chi}^{\left(  2\right)  }=0\ ,\\
g^{\alpha\beta}\Delta_{2;\alpha\beta}^{\left(  2\right)  }=-\frac{i}{2\pi}\ .
\end{array}
\right.
\end{equation}
These two conditions are clearly not compatible with the Schouten identity
which is necessary to generate the second representation $\left[
T_{\sigma\rho}^{AV}\right]  _{2}$ starting from the first one $\left[
T_{\sigma\rho}^{AV}\right]  _{1}$. So, in order to get this wonderful result
it is necessary an illegal trick or to corrupt the mathematics. A consistent
condition would be taken both objects as zero quantities, which gives us%
\begin{equation}
p^{\sigma}T_{\sigma\rho}^{AV}=2m\left[  T_{\rho}^{PV}\right]  +\frac{i}{\pi
}\varepsilon_{\rho\xi}p^{\xi},
\end{equation}
connecting us with the well-known anomalous phenomenon \cite{BERTLMANN-BOOK}
in two-dimensions.

\subsection{$T_{\mu;\sigma\rho}^{AV}$ amplitude}

The second subamplitude of the axial-vector sector is $T_{\mu;\sigma\rho}%
^{AV}$. Through the relationship
\begin{equation}
T_{\mu;\sigma\rho}^{AV}=-\varepsilon_{\sigma\alpha}g^{\alpha\beta}\left[
T_{\mu;\beta\rho}^{VV}\right]  \ ,
\end{equation}
and the result (\ref{res_VV3}) it is straightforward to get%
\begin{align}
T_{\mu;\sigma\rho}^{AV}  &  =\varepsilon_{\sigma\xi}P^{\chi}\left[
\square_{3;\mu\rho\chi\xi}^{\left(  2\right)  }\right]  -\varepsilon
_{\sigma\xi}P_{\rho}\left[  \Delta_{2;\xi\mu}^{\left(  2\right)  }\right]
-\varepsilon_{\sigma\xi}P^{\xi}\left[  \Delta_{2;\rho\mu}^{\left(  2\right)
}\right] \nonumber\\
&  -\varepsilon_{\sigma\xi}P^{\chi}\left\{  g_{\xi\rho}\left[  \Delta
_{2;\mu\chi}^{\left(  2\right)  }\right]  -g_{\mu\rho}\left[  \Delta
_{2;\chi\xi}^{\left(  2\right)  }\right]  -g_{\mu\xi}\left[  \Delta
_{2;\rho\chi}^{\left(  2\right)  }\right]  \right\} \nonumber\\
&  +\frac{p_{\mu}}{2}\left[  T_{\sigma\rho}^{AV}\right]  \ . \label{res_AV3}%
\end{align}
From this result it is expected that it should, by consistency, satisfies four
RAGFs. The two RAGFs obtained for the contractions with $p^{\mu}$ and
$p^{\rho}$ are satisfied without additional hypothesis. In contrast,
contracting (\ref{res_AV3}) with $p^{\sigma}$ and $g^{\mu\sigma}$ gives,
respectively,%
\begin{align}
p^{\sigma}\left[  T_{\mu;\sigma\rho}^{AV}\right]   &  =\left[  T_{\mu;\rho
}^{A}\left(  k_{1}\right)  \right]  -\left[  T_{\mu;\rho}^{A}\left(
k_{2}\right)  \right]  +2m\left[  T_{\mu;\rho}^{PV}\right] \nonumber\\
&  -\varepsilon_{{\rho}\sigma}p^{\sigma}p^{\xi}\left\{  \left[  g^{\alpha
\beta}\square_{3;\alpha\beta\mu\xi}^{\left(  2\right)  }\right]  -g_{\xi\mu
}\left[  g^{\alpha\beta}\Delta_{2;\alpha\beta}^{\left(  2\right)  }\right]
\right\} \nonumber\\
&  -\varepsilon_{\rho\sigma}p^{\sigma}p_{\mu}\left\{  \frac{i}{2\pi}+\left[
g^{\alpha\beta}\Delta_{2;\alpha\beta}^{\left(  2\right)  }\right]  \right\}
\ ,
\end{align}%
\begin{align}
g^{\mu\sigma}\left[  T_{\mu;\sigma\rho}^{AV}\right]   &  =\left[  T_{\rho}%
^{A}\left(  k_{2}\right)  \right]  -m\left[  T_{\rho}^{PV}\right] \nonumber\\
&  -\varepsilon_{\rho\xi}p^{\xi}\left\{  \frac{i}{2\pi}+\left[  g^{\alpha
\beta}\Delta_{2;\alpha\beta}^{\left(  2\right)  }\right]  \right\}  \ .
\end{align}
We see clearly that the RAGFs, generated by such contractions, are not
automatically satisfied. However, if one wants to preserve both of these RAGFs
it is enough to fulfill the conditions (\ref{req_1}) and (\ref{trace_Delta_2})
previously found.

Again, we note that the options $\square_{3;\mu\nu\rho\xi}^{\left(  2\right)
}=\Delta_{2;\mu\nu}^{\left(  2\right)  }=0$ and $g^{\mu\nu}\square_{3;\mu
\nu\rho\xi}^{\left(  2\right)  }=g_{\xi\rho}\left[  g^{\mu\nu}\Delta_{2;\mu
\nu}^{\left(  2\right)  }\right]  =0$ means violations of both RAGF's as
happens in the case of the amplitude $T_{\sigma\rho}^{AV}.$

\subsection{$T_{\mu\nu;\sigma\rho}^{AV}$ Amplitude}

The last subamplitude of this set is $T_{\mu\nu;\sigma\rho}^{AV}$. Its
relationship with $T_{\mu\nu;\beta\rho}^{VV}$, Eq. (\ref{res_VV4}),%
\begin{equation}
T_{\mu\nu;\sigma\rho}^{AV}=-\varepsilon_{\sigma\alpha}g^{\alpha\beta}\left[
T_{\mu\nu;\beta\rho}^{VV}\right]  \ ,
\end{equation}
makes its calculation immediate. The contraction of this result with $p^{\rho
}$ and $p^{\mu}$ reveals that the corresponding RAGFs are satisfied. On the
other hand, the contractions with $p^{\sigma}$ and $g^{\mu\sigma}$ reveal
unexpected (violating) terms given explicitly below%
\begin{align}
p^{\sigma}\left[  T_{{\mu\nu;\sigma\rho}}^{AV}\right]   &  =\left[  T_{\mu
\nu;\rho}^{A}\left(  k_{1}\right)  \right]  -\left[  T_{\mu\nu;\rho}%
^{A}\left(  k_{2}\right)  \right]  +2m\left[  T_{\mu\nu;\rho}^{PV}\right]
\nonumber\\
&  +\frac{1}{3}\varepsilon_{{\rho\sigma}}p^{\sigma}\left(  k_{2}^{\xi}%
k_{2}^{\chi}+k_{1}^{\xi}k_{2}^{\chi}+k_{1}^{\xi}k_{1}^{\chi}\right)  \left[
g^{\alpha\beta}\Sigma_{4;\alpha\beta\mu\nu\xi\chi}^{\left(  2\right)  }\right]
\nonumber\\
&  -\frac{1}{2}\varepsilon_{{\rho}\sigma}p^{\sigma}\left(  k_{1}^{2}+k_{2}%
^{2}\right)  \left[  g^{\alpha\beta}\square_{3;\alpha\beta\mu\nu}^{\left(
2\right)  }\right] \nonumber\\
&  -\varepsilon_{{\rho}\sigma}p^{\sigma}p^{\xi}k_{1{\mu}}\left[
g^{\alpha\beta}\square_{3;\alpha\beta\nu\xi}^{\left(  2\right)  }\right]
-\varepsilon_{{\rho}\sigma}p^{\sigma}p^{\xi}k_{1{\nu}}\left[  g^{\alpha\beta
}\square_{3;\alpha\beta\mu\xi}^{\left(  2\right)  }\right] \nonumber\\
&  +2\varepsilon_{{\rho\sigma}}p^{\sigma}k_{1{\mu}}k_{1{\nu}}\left[
g^{\alpha\beta}\Delta_{2;\alpha\beta}^{\left(  2\right)  }\right]  +\frac
{i}{4\pi}\frac{1}{3}\varepsilon_{\rho\sigma}p^{\sigma}\left[  4p_{\mu}p_{\nu
}-g_{\mu\nu}p^{2}\right]  \ ,
\end{align}%
\begin{align}
g^{\mu\sigma}\left[  T_{{\mu\nu;\sigma\rho}}^{AV}\right]   &  =\left[
T_{\nu;\rho}^{A}\left(  k_{2}\right)  \right]  -m\left[  T_{\nu;\rho}%
^{PV}\right] \nonumber\\
&  +\frac{1}{2}\varepsilon_{{\rho}\xi}p^{\xi}\left(  {k_{2}}+{k_{1}}\right)
^{\sigma}\left[  g^{\alpha\beta}\square_{3;\alpha\beta\nu\sigma}^{\left(
2\right)  }\right] \nonumber\\
&  -\varepsilon_{{\rho}\xi}p^{\xi}k_{1{\nu}}\left[  g^{\alpha\beta}%
\Delta_{2;\alpha\beta}^{\left(  2\right)  }\right]  +\frac{i}{4\pi}%
\varepsilon_{\rho\xi}p^{\xi}p_{\nu}\ .
\end{align}
Entirely similar to what occurred with $T_{{\mu\nu;\sigma\rho}}^{VV}$, in
order to save both RAGFs it is imperative that%
\begin{equation}
\left\{
\begin{array}
[c]{c}%
g^{\mu\nu}\square_{3;\mu\nu\rho\xi}^{\left(  2\right)  }=g_{\xi\rho}\left[
g^{\mu\nu}\Delta_{2;\mu\nu}^{\left(  2\right)  }\right]  =-\frac{i}{2\pi
}g_{\xi\rho}\ ,\\
g^{\mu\sigma}\Sigma_{4;\mu\sigma\nu\rho\xi\chi}^{\left(  2\right)  }=-\frac
{i}{2\pi}\left(  g_{\nu\rho}g_{\xi\chi}+g_{\nu\xi}g_{\rho\chi}+g_{\nu\chi
}g_{\rho\xi}\right)  \ ,
\end{array}
\right.
\end{equation}
which are the same results previously obtained.

We should emphasize, again, that the results above can be easily checked by
assuming the validity of the linearity in the integration operation, as we
have pointed out before. The results for $T_{{\sigma\rho}}^{VA}$,
$T_{{\mu;\sigma\rho}}^{VA}$, and $T_{{\mu\nu;\sigma\rho}}^{VA}$ are completely
analogous to $T_{{\sigma\rho}}^{AV}$, $T_{{\mu;\sigma\rho}}^{AV}$, and
$T_{{\mu\nu;\sigma\rho}}^{AV}$.

\subsection{$T_{\sigma\rho}^{AA}$ Amplitude}

Although the procedure is essentially the same to the one presented above, for
completeness, we succinctly present the main results for the subamplitudes
belonging to the axial sector, obtained through the substitutions: $\Gamma
_{i}=\gamma_{\sigma}\gamma_{3}$ and $\Gamma_{j}=\gamma_{\rho}\gamma_{3}$ in
(\ref{T2}), (\ref{T3}), and (\ref{T4}).

The first subamplitude is $T_{\sigma\rho}^{AA}$, which has the corresponding
result%
\begin{align}
T_{\sigma\rho}^{AA}  &  =2\left[  \Delta_{2;\sigma\rho}^{\left(  2\right)
}\right]  +4\left[  p_{\sigma}p_{\rho}-g_{\sigma\rho}p^{2}\right]  \left[
\xi_{2}^{\left(  -1\right)  }\left(  p^{2},m^{2}\right)  -\xi_{1}^{\left(
-1\right)  }\left(  p^{2},m^{2}\right)  \right] \nonumber\\
&  -8m^{2}g_{\sigma\rho}\left[  \xi_{1}^{\left(  -1\right)  }\left(
p^{2},m^{2}\right)  \right]  . \label{res_AA2}%
\end{align}
Both expected RAGFs (see appendix \ref{Ap_RAGF_Axial_Sec}) are fulfilled by
the above expression without any assumption about the object $\Delta
_{2;\sigma\rho}^{\left(  2\right)  }$.

Again we observe that, similarly to $T_{\sigma\rho}^{VV}$ and $T_{\sigma\rho
}^{AV}$ amplitudes, $T_{\sigma\rho}^{AA}$ belongs to the set of amplitudes
associated with standard (renormalizable) theories. So, within such a context,
this amplitude should satisfy additional constraints such as the two following
WIs%
\begin{equation}
\left\{
\begin{array}
[c]{c}%
p^{\sigma}\left[  T_{\sigma\rho}^{AA}\right]  =2m\left[  T_{\rho}^{PA}\right]
\ ,\\
p^{\rho}\left[  T_{\sigma\rho}^{AA}\right]  =2m\left[  T_{\sigma}^{AP}\right]
\ ,
\end{array}
\right.
\end{equation}
representing the proportionality between the axial current divergence and the
pseudoscalar one, for the case of massive fermions. In order to satisfy both
WIs it is required that%
\begin{equation}
\left[  T_{\rho}^{V}\left(  k_{1}\right)  \right]  -\left[  T_{\rho}%
^{V}\left(  k_{2}\right)  \right]  =0\ ,
\end{equation}
in a completely similar way as in the case of the amplitude $T_{\sigma\rho
}^{VV}$, as expected.

\subsection{$T_{\mu;\sigma\rho}^{AA}$ Amplitude}

The second subamplitude of this set is $T_{\mu;\sigma\rho}^{AA}$ . We found%
\begin{align}
T_{\mu;\sigma\rho}^{AA}  &  =-P^{\alpha}\left[  \square_{3;\alpha\mu\sigma
\rho}^{\left(  2\right)  }\right]  +P_{\rho}\left[  \Delta_{2;\sigma\mu
}^{\left(  2\right)  }\right]  +P_{\sigma}\left[  \Delta_{2;\rho\mu}^{\left(
2\right)  }\right] \nonumber\\
&  +P^{\alpha}\left\{  g_{\sigma\rho}\left[  \Delta_{2;\mu\alpha}^{\left(
2\right)  }\right]  -g_{\mu\rho}\left[  \Delta_{2;\alpha\sigma}^{\left(
2\right)  }\right]  -g_{\mu\sigma}\left[  \Delta_{2;\rho\alpha}^{\left(
2\right)  }\right]  \right\} \nonumber\\
&  -\frac{p_{\mu}}{2}\left[  T_{\sigma\rho}^{AA}\right]  \text{ }.
\label{res_AA3}%
\end{align}
When the RAGFs are checked, we find that those corresponding to the
contractions $p^{\sigma}\left[  T_{\mu;\sigma\rho}^{AA}\right]  $ and $p^{\mu
}\left[  T_{\mu;\sigma\rho}^{AA}\right]  $ (see appendix
\ref{Ap_RAGF_Axial_Sec}) are satisfied, while that related to the contraction
$g^{\mu\sigma}\left[  T_{\mu;\sigma\rho}^{AA}\right]  $ gives%
\begin{align}
g^{\mu\sigma}\left[  T_{\mu;\sigma\rho}^{AA}\right]   &  =\left[  T_{\rho}%
^{V}\left(  k_{2}\right)  \right]  -m\left[  T_{\rho}^{PA}\right] \nonumber\\
&  -\left(  k_{2}+k_{1}\right)  ^{\xi}\left\{  \left[  g^{\alpha\beta}%
\square_{3;\alpha\beta\rho\xi}^{\left(  2\right)  }\right]  +g_{\rho\xi
}\left[  g^{\alpha\beta}\Delta_{2;\alpha\beta}^{\left(  2\right)  }\right]
\right\}  .
\end{align}
The conclusions are the same ones obtained for $T_{\mu;\sigma\rho}^{VV}$.

\subsection{$T_{\mu\nu;\sigma\rho}^{AA}$ Amplitude}

The last one to be calculated is $T_{\mu\nu;\sigma\rho}^{AA}$ . The result can
be put in the form%
\begin{align}
T_{\sigma\rho;\mu\nu}^{AA}  &  =S_{\sigma\rho;\mu\nu}^{VV}+\left(  g_{\rho\mu
}g_{\sigma\nu}+g_{\rho\nu}g_{\sigma\mu}\right)  \left[  I_{quad}^{\left(
2\right)  }\left(  m^{2}\right)  \right] \nonumber\\
&  +\left\{  \frac{1}{3}\left(  g_{\mu\nu}g_{\rho\sigma}p^{2}-g_{\mu\nu
}p_{\rho}p_{\sigma}-g_{\sigma\rho}p_{\mu}p_{\nu}\right)  \right. \nonumber\\
&  \text{ \ \ \ \ \ }-\frac{1}{6}\left(  g_{\mu\rho}g_{\nu\sigma}p^{2}%
-g_{\mu\rho}p_{\nu}p_{\sigma}-g_{\nu\sigma}p_{\mu}p_{\rho}\right) \nonumber\\
&  \text{ \ \ \ \ }\left.  -\frac{1}{6}\left(  g_{\nu\rho}g_{\mu\sigma}%
p^{2}-g_{\nu\rho}p_{\mu}p_{\sigma}-g_{\mu\sigma}p_{\nu}p_{\rho}\right)
-2m^{2}g_{\mu\nu}g_{\sigma\rho}\right\}  \left[  I_{\log}^{\left(  2\right)
}\left(  m^{2}\right)  \right] \nonumber\\
&  +\frac{i}{2\pi}\left\{  \frac{p_{\mu}p_{\nu}p_{{\sigma}}p_{\rho}}{p^{2}%
}+\frac{p^{2}}{2}\left(  g_{\nu{\rho}}g_{\mu{\sigma}}+g_{\nu\sigma}g_{\mu
{\rho}}\right)  \right. \nonumber\\
&  \ \ \ \ \ \ \ \ \ \ \ \ \left.  -\frac{p_{\nu}}{2}\left(  g_{\mu\sigma
}p_{\rho}+g_{\mu\rho}p_{\sigma}\right)  -\frac{p_{\mu}}{2}\left(  g_{\nu{\rho
}}p_{{\sigma}}+g_{\nu{\sigma}}p_{{\rho}}\right)  \right\}  \left[  2\xi
_{2}^{\left(  0\right)  }-\xi_{1}^{\left(  0\right)  }\right] \nonumber\\
&  +\frac{i}{2\pi}\frac{1}{p^{2}}\left(  p_{\sigma}p_{\rho}-p^{2}%
g_{{\sigma\rho}}\right)  \left(  p_{\mu}p_{\nu}-p^{2}g_{\mu\nu}\right)
\left[  \xi_{2}^{\left(  0\right)  }-\xi_{1}^{\left(  0\right)  }\right]
\nonumber\\
&  -\frac{i}{2\pi}\frac{m^{2}}{p^{2}}g_{\mu\nu}\left(  p_{\sigma}p_{\rho
}-p^{2}g_{\sigma\rho}\right)  \left[  \xi_{0}^{\left(  0\right)  }\right]
+\frac{1}{4}p_{\rho}p_{\sigma}\left[  T_{\mu\nu}^{AA}\right]  \ .
\label{res_AA4}%
\end{align}
The verification of the RAGFs shows that only the contraction $g^{\mu\sigma
}T_{\mu\nu;\sigma\rho}^{AA}$ is not automatically satisfied. Instead, we find%
\begin{align}
g^{\mu\sigma}T_{\mu\nu;\sigma\rho}^{AA}  &  =\left[  T_{\nu;\rho}^{V}\left(
k_{2}\right)  \right]  -m\left[  T_{\nu;\rho}^{PA}\right] \nonumber\\
&  +\frac{1}{3}\left(  k_{2}^{\xi}k_{2}^{\chi}+k_{1}^{\xi}k_{2}^{\chi}%
+k_{1}^{\xi}k_{1}^{\chi}\right)  \left[  g^{\mu\sigma}\Sigma_{4;\mu\sigma
\nu\rho\xi\chi}^{\left(  2\right)  }\right] \nonumber\\
&  -\frac{1}{2}\left(  k_{1}^{2}+k_{2}^{2}\right)  \left[  g^{\mu\sigma
}\square_{3;\mu\sigma\nu\rho}^{\left(  2\right)  }\right] \nonumber\\
&  -\frac{1}{2}\left(  k_{2}+k_{1}\right)  _{\rho}\left(  k_{2}+k_{1}\right)
^{\xi}\left[  g^{\mu\sigma}\square_{3;\mu\sigma\nu\xi}^{\left(  2\right)
}\right] \nonumber\\
&  -k_{1\nu}\left(  k_{2}+k_{1}\right)  ^{\xi}\left[  g^{\mu\sigma}%
\square_{3;\mu\sigma\rho\xi}^{\left(  2\right)  }\right] \nonumber\\
&  +k_{1\nu}\left(  k_{2}+k_{1}\right)  _{\rho}\left[  g^{\mu\sigma}%
\Delta_{2;\mu\sigma}^{\left(  2\right)  }\right] \nonumber\\
&  +\frac{1}{3}\left[  p_{\nu}p_{\rho}-g_{\nu\rho}p^{2}\right]  \ .
\end{align}
Clearly, the conditions that ensure this RAGFs (see Eq.
(\ref{RAGF_AA_4_index_g})) be fulfilled are the same ones required for the
$T_{\mu\nu;\sigma\rho}^{VV}$ and $T_{\mu\nu;\sigma\rho}^{AV}$, as discussed above.

\section{RAGFs versus Einstein and Weyl Gravitational Anomalies}

In the section \ref{sec_exp_eval} we shown that, in order to satisfy
(simultaneously) all the RAGFs expected for the subamplitudes of $T_{\mu
\nu\rho\sigma}^{\left(  G\right)  }$, it is required a set of conditions
involving finite quantities. They represent necessary conditions for the
calculation procedure be consistent with the linearity operation in the
integrals. At this point we can ask ourselves: are the above requirements
enough to guarantee also the maintenance of the RAGFs associated with
$T_{\mu\nu\rho\sigma}^{\left(  G\right)  }$? Given the investigation about the
subamplitudes made in the previous section, the answer to this query is
immediate. This is because the gravitational amplitude $T_{\mu\nu\rho\sigma
}^{\left(  G\right)  }$ was decomposed into a sum of subamplitudes. Thus, when
we contract $T_{\mu\nu\rho\sigma}^{\left(  G\right)  }$ with $p^{\mu}$ or
$g^{\mu\nu}$, we get contractions with these subamplitudes with $p^{\mu}$ or
$g^{\mu\nu}$ also, and, each contraction generates a RAGFs for such
subamplitudes, as we saw in the section \ref{sec_exp_eval}. For instance, in
the vector sector we have%
\begin{align}
p^{\mu}T_{\mu\nu\rho\sigma}^{\left(  V\right)  }  &  =4\left\{  \left[
p^{\mu}T_{\nu\sigma;\mu\rho}^{VV}\right]  +\left[  p^{\mu}T_{\mu\sigma;\nu
\rho}^{VV}\right]  \right\} \nonumber\\
&  +2p_{\sigma}\left\{  \left[  p^{\mu}T_{\nu;\mu\rho}^{VV}\right]  +\left[
p^{\mu}T_{\mu;\nu\rho}^{VV}\right]  \right\} \nonumber\\
&  +2p_{\nu}\left[  p^{\mu}T_{\sigma;\mu\rho}^{VV}\right]  +p_{\nu}p_{\sigma
}\left[  p^{\mu}T_{\mu\rho}^{VV}\right] \nonumber\\
&  +2p^{2}\left[  T_{\sigma;\nu\rho}^{VV}\right]  +p_{\sigma}p^{2}\left[
T_{\nu\rho}^{VV}\right] \nonumber\\
&  +\left(  \sigma\longleftrightarrow\rho\right)  \ ,
\end{align}
and%
\begin{align}
g^{\mu\nu}\left[  T_{\mu\nu\rho\sigma}^{\left(  V\right)  }\right]   &
=8\left[  g^{\mu\nu}T_{\mu\sigma;\nu\rho}^{VV}\right]  +4p^{\sigma}\left[
g^{\mu\nu}T_{\mu;\nu\rho}^{VV}\right] \nonumber\\
&  +4\left[  p^{\mu}T_{\sigma;\mu\rho}^{VV}\right]  +2p_{\sigma}\left[
p^{\mu}T_{\mu\rho}^{VV}\right] \nonumber\\
&  +\left(  \sigma\leftrightarrow\rho\right)  \text{ }.
\end{align}
So, obviously, if all subamplitudes fulfill its RAGFs then $T_{\mu\nu
\rho\sigma}^{\left(  G\right)  }$ fulfill its RAGFs as well. So far so good.
However, this game becomes more complex when WIs are expected to be preserved too.

\subsection{Ward Identities and Gravitational Anomalies}

As we have argued along the work, the preservation of the RAGFs can be
considered as a requirement of purely mathematical nature. From a physical
point of view, the main question to be considered is about the WIs. Both
aspects, however, seems to be coupled. Precisely due to this reason we
reported the investigation in the way presented previously. As a summary, we
saw that the (regularization independent) conditions for the RAGFs
preservation for all subamplitudes are%
\begin{equation}
\left\{
\begin{array}
[c]{c}%
g^{\mu\nu}\square_{3;\mu\nu\rho\xi}^{\left(  2\right)  }=g_{\xi\rho}\left[
g^{\mu\nu}\Delta_{2;\mu\nu}^{\left(  2\right)  }\right]  =-\frac{i}{2\pi
}g_{\xi\rho}\ ,\\
g^{\mu\nu}\Sigma_{4;\mu\nu\sigma\rho\xi\chi}^{\left(  2\right)  }=-\frac
{i}{2\pi}\left(  g_{\sigma\rho}g_{\xi\chi}+g_{\sigma\xi}g_{\rho\chi}%
+g_{\sigma\chi}g_{\rho\xi}\right)  \ ,
\end{array}
\right.
\end{equation}
while the (regularization dependent) conditions for the WIs maintenance, in
the two-Lorentz index amplitudes, are%
\begin{equation}
\Sigma_{4;\mu\nu\sigma\rho\xi\chi}^{\left(  2\right)  }=\square_{3;\mu\nu
\rho\xi}^{\left(  2\right)  }=\Delta_{2;\mu\nu}^{\left(  2\right)  }=0\ .
\end{equation}
The issue is that these conditions are conflicting ones.

With a few exceptions, all these calculations are usually performed after a
regularization prescription is adopted. Within this context, in the most of
time, it is not possible to see clearly that, after all calculations, the main
difference among two distinct approaches resides on the value attributed to
the aforementioned objects. Note that it is not relevant the way we write the
scalar divergent objects. Relative to the possible values that can be
attributed to the surface terms, it is reasonable to consider three different
scenarios, which we will discuss in details in what follows.

In a first scenario one can adopt a prescription where all the surface terms
defined above are null tensors, i.e.%
\begin{equation}
\Sigma_{4;\mu\nu\sigma\rho\xi\chi}^{\left(  2\right)  }=\square_{3;\mu\nu
\rho\xi}^{\left(  2\right)  }=\square_{2;\mu\nu\rho\xi}^{\left(  2\right)
}=\Delta_{2;\mu\nu}^{\left(  2\right)  }=\Delta_{1;\mu\nu}^{\left(  2\right)
}=0\ , \label{cons_rel}%
\end{equation}
as well as their contractions with $g^{\mu\nu}$ (tensor traces)
\begin{equation}
g^{\mu\nu}\Sigma_{4;\mu\nu\sigma\rho\xi\chi}^{\left(  2\right)  }=g^{\mu\nu
}\square_{3;\mu\nu\rho\xi}^{\left(  2\right)  }=g^{\mu\nu}\square_{2;\mu
\nu\rho\xi}^{\left(  2\right)  }=g^{\mu\nu}\Delta_{2;\mu\nu}^{\left(
2\right)  }=g^{\mu\nu}\Delta_{1;\mu\nu}^{\left(  2\right)  }=0\ .
\end{equation}
These assumptions are consistent, in a trivial way, for example, with the
Schouten identity
\begin{equation}
\varepsilon_{\sigma\xi}g^{\xi\chi}\left[  \Delta_{2;\chi\rho}^{\left(
2\right)  }\right]  -\varepsilon_{\rho\xi}g^{\xi\chi}\left[  \Delta
_{2;\chi\sigma}^{\left(  2\right)  }\right]  =\varepsilon_{\sigma\rho}\left[
g^{\xi\chi}\Delta_{2;\xi\chi}^{\left(  2\right)  }\right]  \ ,
\end{equation}
the one that was required to verify whether the RAGFs for the pseudoamplitudes
are preserved by the calculations made. As a consequence of these assumptions,
all the one-point functions vanish identically, at least in the massless
limit, which are physical desirable results. On the other side, they are not
consistent with the linearity in the integration operation, as we have seen.
We can look at properties (\ref{cons_rel}) as a kind of (physical) consistency
relations. The authors of the present work have shown, in different contexts
involving perturbative calculations
\cite{ORIMAR-PRD1,ORIMAR-PRD2,ORIMAR-NP,ORIMAR-NJL} that the consistency
relations are required in order to preserve gauge invariance as well as to
eliminate ambiguous terms, as we have shown for the two-Lorentz index
amplitudes in the section (\ref{sec_exp_eval}). In general, the conditions
(\ref{cons_rel}) are satisfied by a class of regularizations called gauge
preserving regularizations, of which DR is the most popular member.

By assuming the conditions above we can define what we may call the
\textquotedblleft physical\textquotedblright\ (sub)amplitudes:%
\begin{equation}
\mathcal{T}_{\sigma\rho}^{VV}=\frac{i}{\pi}\left(  p_{\sigma}p_{\rho
}-g_{\sigma\rho}p^{2}\right)  \left[  \xi_{2}^{\left(  -1\right)  }-\xi
_{1}^{\left(  -1\right)  }\right]  \ ,
\end{equation}%
\begin{equation}
\mathcal{T}_{\mu;\sigma\rho}^{VV}=-\frac{p_{\mu}}{2}\left[  \mathcal{T}%
_{\sigma\rho}^{VV}\right]  \ ,
\end{equation}%
\begin{align}
\mathcal{T}_{\mu\nu;\sigma\rho}^{VV}  &  =\left(  g_{\nu\sigma}g_{\mu\rho
}+g_{\nu\rho}g_{\mu\sigma}\right)  \left[  I_{quad}^{\left(  2\right)
}\left(  m^{2}\right)  \right] \nonumber\\
&  +\left\{  \frac{1}{3}\left(  g_{\sigma\rho}g_{\nu\mu}p^{2}-g_{\sigma\rho
}p_{\nu}p_{\mu}-g_{\mu\nu}p_{\sigma}p_{\rho}\right)  \right. \nonumber\\
&  \text{ \ }\ \ \ \ -\frac{1}{6}\left(  g_{\sigma\nu}g_{\rho\mu}%
p^{2}-g_{\sigma\nu}p_{\rho}p_{\mu}-g_{\rho\mu}p_{\sigma}p_{\nu}\right)
\nonumber\\
&  \text{ \ \ \ \ }\left.  -\frac{1}{6}\left(  g_{\rho\nu}g_{\sigma\mu}%
p^{2}-g_{\rho\nu}p_{\sigma}p_{\mu}-g_{\sigma\mu}p_{\rho}p_{\nu}\right)
\right\}  \left[  I_{\log}^{\left(  2\right)  }\left(  m^{2}\right)  \right]
\nonumber\\
&  +\frac{i}{2\pi}\left\{  \frac{p_{\sigma}p_{\rho}p_{{\mu}}p_{\nu}}{p^{2}%
}+\frac{p^{2}}{2}\left(  g_{\rho{\nu}}g_{\sigma{\mu}}+g_{\rho\mu}g_{\sigma
{\nu}}\right)  \right. \nonumber\\
&  \text{ \ \ \ \ \ \ \ \ \ \ \ }\left.  -\frac{p_{\rho}}{2}\left(
g_{\sigma\mu}p_{\nu}+g_{\sigma\nu}p_{\mu}\right)  -\frac{p_{\sigma}}{2}\left(
g_{\rho{\nu}}p_{{\mu}}+g_{\rho{\mu}}p_{{\nu}}\right)  \right\}  \left[
2\xi_{2}^{\left(  0\right)  }-\xi_{1}^{\left(  0\right)  }\right] \nonumber\\
&  +\frac{i}{2\pi}\frac{1}{p^{2}}\left(  p_{\mu}p_{\nu}-p^{2}g_{{\mu\nu}%
}\right)  \left(  p_{\sigma}p_{\rho}-p^{2}g_{\sigma\rho}\right)  \left[
\xi_{2}^{\left(  0\right)  }-\xi_{1}^{\left(  0\right)  }\right]  +\frac{1}%
{4}p_{\nu}p_{\mu}\left[  \mathcal{T}_{\sigma\rho}^{VV}\right]  \ ,
\end{align}%
\begin{align}
\mathcal{T}_{\mu\nu}^{AV}  &  =-\varepsilon_{\mu\alpha}g^{\alpha\beta}\left[
\mathcal{T}_{\beta\nu}^{VV}\right]  \ ,\\
\mathcal{T}_{\mu;\sigma\rho}^{AV}  &  =-\varepsilon_{\sigma\alpha}%
g^{\alpha\beta}\left[  \mathcal{T}_{\mu;\beta\rho}^{VV}\right]  \ ,\\
\mathcal{T}_{\mu\nu;\sigma\rho}^{AV}  &  =-\varepsilon_{\sigma\alpha}%
g^{\alpha\beta}\left[  \mathcal{T}_{\mu\nu;\beta\rho}^{VV}\right]  \ ,
\end{align}%
\begin{equation}
\mathcal{T}_{\sigma\rho}^{AA}=\frac{i}{\pi}\left\{  \left(  p_{\sigma}p_{\rho
}-g_{\sigma\rho}p^{2}\right)  \left[  \xi_{2}^{\left(  -1\right)  }-\xi
_{1}^{\left(  -1\right)  }\right]  -g_{\sigma\rho}m^{2}\left[  \xi
_{0}^{\left(  -1\right)  }\right]  \right\}  \ ,\nonumber
\end{equation}%
\begin{equation}
\mathcal{T}_{\mu;\sigma\rho}^{AA}=-\frac{p_{\mu}}{2}\left[  \mathcal{T}%
_{\sigma\rho}^{AA}\right]  \text{ },
\end{equation}%
\begin{align}
\mathcal{T}_{\mu\nu;\sigma\rho}^{AA}  &  =\left(  g_{\nu\sigma}g_{\mu\rho
}+g_{\nu\rho}g_{\mu\sigma}\right)  \left[  I_{quad}^{\left(  2\right)
}\left(  m^{2}\right)  \right] \nonumber\\
&  +\left\{  \frac{1}{3}\left(  g_{\sigma\rho}g_{\nu\mu}p^{2}-g_{\sigma\rho
}p_{\nu}p_{\mu}-g_{\mu\nu}p_{\sigma}p_{\rho}\right)  \right. \nonumber\\
&  \text{ \ \ \ \ \ }-\frac{1}{6}\left(  g_{\sigma\nu}g_{\rho\mu}%
p^{2}-g_{\sigma\nu}p_{\rho}p_{\mu}-g_{\rho\mu}p_{\sigma}p_{\nu}\right)
\nonumber\\
&  \text{ \ \ \ \ }\left.  -\frac{1}{6}\left(  g_{\rho\nu}g_{\sigma\mu}%
p^{2}-g_{\rho\nu}p_{\sigma}p_{\mu}-g_{\sigma\mu}p_{\rho}p_{\nu}\right)
-2m^{2}g_{\sigma\rho}g_{\mu\nu}\right\}  \left[  I_{\log}^{\left(  2\right)
}\left(  m^{2}\right)  \right] \nonumber\\
&  +\frac{i}{2\pi}\left\{  \frac{p_{\sigma}p_{\rho}p_{{\mu}}p_{\nu}}{p^{2}%
}+\frac{p^{2}}{2}\left(  g_{\rho{\nu}}g_{\sigma{\mu}}+g_{\rho\mu}g_{\sigma
{\nu}}\right)  \right. \nonumber\\
&  \ \ \ \ \ \ \ \ \ \ \ \ \left.  -\frac{p_{\rho}}{2}\left(  g_{\sigma\mu
}p_{\nu}+g_{\sigma\nu}p_{\mu}\right)  -\frac{p_{\sigma}}{2}\left(  g_{\rho
{\nu}}p_{{\mu}}+g_{\rho{\mu}}p_{{\nu}}\right)  \right\}  \left[  2\xi
_{2}^{\left(  0\right)  }-\xi_{1}^{\left(  0\right)  }\right] \nonumber\\
&  +\frac{i}{2\pi}\frac{1}{p^{2}}\left(  p_{\mu}p_{\nu}-p^{2}g_{{\mu\nu}%
}\right)  \left(  p_{\sigma}p_{\rho}-p^{2}g_{\sigma\rho}\right)  \left[
\xi_{2}^{\left(  0\right)  }-\xi_{1}^{\left(  0\right)  }\right] \nonumber\\
&  -\frac{i}{2\pi}\frac{m^{2}}{p^{2}}g_{\sigma\rho}\left(  p_{\mu}p_{\nu
}-p^{2}g_{\mu\nu}\right)  \left[  \xi_{0}^{\left(  0\right)  }\right]
+\frac{1}{4}p_{\mu}p_{\nu}\left[  \mathcal{T}_{\sigma\rho}^{AA}\right]  \ .
\end{align}
Observe that all ambiguous terms were eliminated and all the one-point
functions vanished (for massless case). From these redefined subamplitudes we
get%
\begin{align}
p^{\mu}\mathcal{T}_{\mu\nu\rho\sigma}^{G}  &  =-4\left(  p_{\sigma}g_{\rho\nu
}+2p_{\nu}g_{\rho\sigma}+p_{\rho}g_{\nu\sigma}\right)  \left[  I_{quad}%
^{\left(  2\right)  }\left(  m^{2}\right)  \right] \nonumber\\
&  +2m^{2}\left\{  p_{\rho}g_{\nu\sigma}\left[  I_{\log}^{\left(  2\right)
}\left(  m^{2}\right)  \right]  +p_{\sigma}g_{\nu\rho}\left[  I_{\log
}^{\left(  2\right)  }\left(  m^{2}\right)  \right]  \right. \nonumber\\
&  \ \ \ \ \ \ \ \ \ \ +g_{\nu\rho}p^{\alpha}g_{\sigma\alpha}\left[  I_{\log
}^{\left(  2\right)  }\left(  m^{2}\right)  \right]  +g_{\nu\sigma}p^{\alpha
}g_{\rho\alpha}\left[  I_{\log}^{\left(  2\right)  }\left(  m^{2}\right)
\right] \nonumber\\
&  \ \ \ \ \ \ \ \ \ \left.  +\frac{1}{64\pi}\left[  p_{\rho}\left(  p_{\nu
}p_{\sigma}-g_{\nu\sigma}p^{2}\right)  +p_{\sigma}\left(  p_{\nu}p_{\rho
}-g_{\nu\rho}p^{2}\right)  \pm2\varepsilon_{\beta\nu}p_{\beta}\left(
p_{\sigma}p_{\rho}-g_{\sigma\rho}p^{2}\right)  \right]  \left[  \xi
_{1}^{\left(  -1\right)  }-2\xi_{2}^{\left(  -1\right)  }\right]  \right\}
\nonumber\\
&  \mp\frac{1}{96\pi}\varepsilon_{\beta\nu}p_{\beta}\left(  p_{\sigma}p_{\rho
}-g_{\sigma\rho}p^{2}\right)  \ ,
\end{align}
or, by taking the massless limit%
\begin{equation}
p^{\mu}\mathcal{T}_{\mu\nu\rho\sigma}^{G}=\mp\frac{1}{96\pi}\varepsilon
_{\beta\nu}p^{\beta}\left(  p_{\sigma}p_{\rho}-g_{\sigma\rho}p^{2}\right)  \ ,
\end{equation}
which can be recognized as being the well-known Einstein's gravitational
anomaly \cite{BERTLMANN-2}.

In the same way, the $g^{\mu\nu}T_{\mu\nu\rho\sigma}^{G}$ contraction gives%
\begin{align}
g^{\mu\nu}\mathcal{T}_{\mu\nu\rho\sigma}^{G}  &  =4g_{\sigma\rho}\left\{
3\left[  I_{quad}^{\left(  2\right)  }\left(  m^{2}\right)  \right]
-2m^{2}\left[  I_{\log}^{\left(  2\right)  }\left(  m^{2}\right)  \right]
\right\} \nonumber\\
&  -\left(  \frac{1}{8\pi}\right)  m^{2}\left[  \left(  p_{\sigma}p_{\rho
}-p^{2}g_{\rho\sigma}\right)  \pm\frac{1}{2}\left(  \varepsilon_{\mu\rho
}p^{\mu}p_{\sigma}+\varepsilon_{\mu\sigma}p^{\mu}p_{\rho}\right)  \right]
\left[  \xi_{1}^{\left(  -1\right)  }-2\xi_{2}^{\left(  -1\right)  }\right]
\nonumber\\
&  +\left(  \frac{1}{24\pi}\right)  \left[  \left(  p_{\sigma}p_{\rho}%
-p^{2}g_{\rho\sigma}\right)  \mp\frac{1}{4}\left(  \varepsilon_{\rho\lambda
}p^{\lambda}p_{\sigma}+\varepsilon_{\sigma\lambda}p^{\lambda}p_{\rho}\right)
\right]  \ ,
\end{align}
or
\begin{equation}
g^{\mu\nu}\mathcal{T}_{\mu\nu\rho\sigma}^{G}=\left(  \frac{1}{24\pi}\right)
\left[  \left(  p_{\sigma}p_{\rho}-p^{2}g_{\rho\sigma}\right)  \mp\frac{1}%
{4}\left(  \varepsilon_{\rho\lambda}p^{\lambda}p_{\sigma}+\varepsilon
_{\sigma\lambda}p^{\lambda}p_{\rho}\right)  \right]  \ ,
\end{equation}
for a massless fermion. This result is known as Weyl or trace gravitational
anomaly. The main point of the preceding calculation is the demonstration
that, from our general results shown in the section (\ref{sec_exp_eval}), one
can obtain the usual anomalies terms. The caveat, as should be clear, is that
both anomalies are inevitably entangled to a violation of a basic mathematical
property, the linearity in the integration operation. This aspect is, in fact,
common to all anomaly phenomena in QFT.

A second possible track that one may follows, if a regularization get into the
game, is to adopt a procedure where surface terms are taken as null objects,
as above, but instead their contractions with $g^{\mu\nu}$
\begin{equation}
\left\{
\begin{array}
[c]{c}%
g^{\mu\nu}\square_{3;\mu\nu\rho\xi}^{\left(  2\right)  }=g_{\xi\rho}\left[
g^{\mu\nu}\Delta_{2;\mu\nu}^{\left(  2\right)  }\right]  =-\frac{i}{2\pi
}g_{\xi\rho}\ ,\\
g^{\mu\sigma}\Sigma_{4;\mu\sigma\nu\rho\xi\chi}^{\left(  2\right)  }=-\frac
{i}{2\pi}\left(  g_{\nu\rho}g_{\xi\chi}+g_{\nu\xi}g_{\rho\chi}+g_{\nu\chi
}g_{\rho\xi}\right)  \ ,
\end{array}
\right.  \label{ST_contractions}%
\end{equation}
being, in effect, non-null quantities. This situation may occurs indirectly
depending on the step of calculations that a regularization is implemented.
For instance, when one takes the surface terms as being zero after the use of
the identity (\ref{Schouten}), this choice can be materialized. Then, within
this paradoxical scenario, through a specific route followed in the
calculations, it would be possible to get the linearity in the integration
operation maintained and so no anomalous terms will survive. It seems that the
best of the outcomes is achieved. One could propose a rule for the evaluation
of perturbative amplitudes: for such referred amplitudes, when convenient,
first use the identity (\ref{Schouten}) and calculate the quantities
$g^{\mu\nu}\Delta_{2;\mu\nu}^{\left(  2\right)  }$, $g^{\mu\nu}\square
_{3;\mu\nu\rho\xi}^{\left(  2\right)  }$, $g^{\mu\sigma}\Sigma_{4;\mu\sigma
\nu\rho\xi\chi}^{\left(  2\right)  }$ obtaining a non-zero value. After that,
take the surface terms as being zero. The desirable results seems to be
obtained and, at first sight, one can understand that, with this recipe, the
anomalies are eliminated since all WIs can be fulfilled. However, if we use
the Schouten identity, at least one of the expressions for the contracted
tensors will contain terms like $g^{\mu\nu}\Delta_{2;\mu\nu}^{\left(
2\right)  }$. This situation reflects the mathematical impossibility of
satisfying all the RAGFs without to use a Schouten identity like
(\ref{Schouten}). So, the situation above, where all the WIs are preserved
cannot, in fact, occur. In addition, \ the Schouten identity is violated since
the tensors are null quantities and their traces are not.

A third possibility would be a choice where both, the surface terms as well as
their traces are non-null tensors. For instance, if a regularization gives
$\left.  \Delta_{2;\mu\nu}^{\left(  2\right)  }\right\vert _{reg}=-\frac
{i}{4\pi}g_{\mu\nu}$, then its trace is given by $g^{\mu\nu}\Delta_{2;\mu\nu
}^{\left(  2\right)  }=-\frac{i}{2\pi}$, the same value we found before. These
assumptions are consistent with the preservation of the linearity in the
integration operation and, consequently, with the uniqueness of the results
since the Schouten identity is preserved also. One can say that this attitude
represents the mathematical consistency. On the other hand, in this case the
one-point functions are nonzero and ambiguous. This is, of course, undesirable
just because the physical amplitudes are ambiguous. The WIs, therefore, are
not preserved due to ambiguous terms. Of course, this class of regularizations
yields a scenario which is not useful to make physical predictions in spite of
being mathematically consistent. If we follow this path, some kind of
procedure must be adopted, as an additional ingredient to the Feynman rules,
in order to eliminate the ambiguous quantities arising.

\section{Summary and Conclusions}

Across this paper, we have calculated the perturbative gravitational amplitude
$T_{\mu\nu\rho\sigma}^{\left(  G\right)  }$. This amplitude was constructed
through Feynman rules derived from a two-dimensional interaction Lagrangian
where Weyl fermions couple to the gravitational field via the energy-momentum
tensor. To organize the intermediate calculations and emphasize key aspects of
the analysis, we decomposed the $T_{\mu\nu\rho\sigma}^{\left(  G\right)  }$
amplitude into sets of subamplitudes based on their tensor character. Each
subamplitude was analyzed using a novel method designed to handle divergent
Feynman integrals without limitations and treat both tensors and pseudotensors
equally. Crucially, this method is not a regularization scheme, as no
divergent integrals are calculated during intermediate steps. Instead, the
undefined content of each amplitude is isolated and expressed as either
surface terms or scalar objects devoid of physical parameters. For amplitudes
with linear or higher divergence, the surface term coefficients capture all
ambiguities arising from the chosen internal loop momenta. Conversely, the
finite content is integrated directly and organized into convenient functions
represented by Feynman parameter integrals. Section \ref{sec_exp_eval}
explicitly demonstrates these features for the calculated gravitational subamplitudes.

A fundamental question arose after applying the proposed method and obtaining
the results: do they satisfy the expected RAGFs? This is crucial because
failing to satisfy the RAGFs indicates a breaking of linearity in the
integration operation, rendering the results unacceptable. For all
subamplitudes to simultaneously satisfy their expected RAGFs, a set of
conditions is necessary. These conditions involve finite quantities,
interpretable as traces over objects identified as surface terms
(\ref{ST_contractions}). Our direct calculations revealed that these
conditions are universally satisfied, independent of the chosen prescription.
This signifies that our procedure, despite dealing with undefined quantities,
yields results consistent with the required linearity of integration. As such,
our calculations align with the established mathematical requirements for the
considered perturbative amplitudes.

The discussion now shifts to the physical interpretation of the results. While
ensuring linearity in integration is crucial for mathematical consistency,
satisfying WIs is essential for interpreting predictions as consequences of
the theory's symmetries. For a universal analysis of perturbative amplitudes,
the calculation method should be independent of the specific theory they
originate from. Within the gravitational amplitude $T_{\mu\nu\rho\sigma
}^{\left(  G\right)  }$, four subamplitudes $T_{\sigma\rho}^{VV}$,
$T_{\sigma\rho}^{AV}$, $T_{\sigma\rho}^{VA}$, and $T_{\sigma\rho}^{AA}$ are
constrained to fulfill WIs if considered part of a gauge theory with conserved
vector currents (e.g., $QED_{2}$). We found that satisfying all WIs for
$T_{\sigma\rho}^{VV}$ and $T_{\sigma\rho}^{AA}$ requires $\Delta_{2;\xi\chi
}^{\left(  2\right)  }=0$. This condition also preserves the vector WI for
$T_{\sigma\rho}^{AV}$ but violates the axial WI, exhibiting the expected $2D$
anomaly. However, setting $\Delta_{2;\xi\chi}^{\left(  2\right)  }=0$ assigns
a defined value to an undefined object, potentially through a regularization
scheme. This directly contradicts the condition $g^{\mu\nu}\Delta_{2;\mu\nu
}^{\left(  2\right)  }=-\frac{i}{2\pi}$, necessary for preserving RAGFs
involving pseudotensor amplitudes. This contradiction arises if $g^{\mu\nu
}\Delta_{2;\mu\nu}^{\left(  2\right)  }$ is interpreted as the trace of
$\Delta_{2;\mu\nu}^{\left(  2\right)  }$, implying a null tensor with a
non-null trace. Additionally, this pair of values violates the Schouten
identity. Even for this relatively simple problem, this scenario presents
significant difficulties.

We observed that something similar occurred with the gravitational amplitude.
We investigate three distinct scenarios corresponding to different choices for
surface terms, recognizing that these choices effectively represent the
selection of specific regularization schemes. In the first scenario, we made
the assumption that there exists a procedure or regularization scheme capable
of assigning a null value to the surface terms and their contractions with the
metric. Consequently, all one-point functions vanish, and ambiguous terms
connected to internal lines momenta are eliminated. This is because ambiguous
combinations of internal lines momenta always act as coefficients of surface
terms. Crucially, this approach preserves Schouten identities that involve
surface terms, such as,%
\begin{equation}
\varepsilon_{\sigma\xi}g^{\xi\chi}\left[  \Delta_{2;\chi\rho}^{\left(
2\right)  }\right]  -\varepsilon_{\rho\xi}g^{\xi\chi}\left[  \Delta
_{2;\chi\sigma}^{\left(  2\right)  }\right]  =\varepsilon_{\sigma\rho}\left[
g^{\xi\chi}\Delta_{2;\xi\chi}^{\left(  2\right)  }\right]  \ ,
\end{equation}
ensuring compatibility with the RAGFs of pseudoamplitudes. Interestingly, this
scenario aligns with the previously analyzed anomalous $T_{\sigma\rho}^{AV}$
amplitude. Within this context, the customary gravitational anomalies are
recovered. From a physical perspective, it appears that these choices are
suitable, demonstrating that our procedure can, once again, replicate the
conventional results obtained through other techniques. However, from a purely
mathematical standpoint, it introduces a contradiction. Presuming the
contractions of surface terms to be null implicitly violates the principle of
linearity in integration, which should hold true even for divergent integrals.
While physically appealing, adopting null surface term contractions leads to a
mathematically inconsistent outcome, unveiling a potential drawback associated
with this particular choice.

Without questioning consistency, we supposed a second scenario that differs
from the previous one by assuming non-null contractions of the surface terms
with the metric (\ref{ST_contractions}), while retaining the assumption of
null surface terms themselves. This choice, validated through simple
calculations where the principle of linearity in integration holds, might
seemingly allow satisfying all WIs for the $T_{\mu\nu\rho\sigma}^{\left(
G\right)  }$ amplitude. However, this apparent solution presents a fundamental
obstacle. The Schouten identity, crucial for maintaining these WIs, becomes
violated under this scenario. This violation exposes the inherent
incompatibility of this choice with mathematical consistency, rendering the
seemingly attainable solution physically implausible.

As the final possibility, we explore a scenario where neither the surface
terms nor their contractions with the metric are assumed to be zero. This
approach upholds the linearity of the integration operation, guaranteeing its
mathematical consistency in this regard. However, this path comes at a cost.
The resulting amplitudes exhibit broken symmetry relations and remain
ambiguous quantities. This compromise in physical interpretation renders the
obtained results unsuitable for predictive purposes.

This exploration of perturbative gravitational anomalies within a simple 2D
model unveils some crucial insights regarding the limitations of
regularizations. Notably, the results obtained are not unique, merely
representing one possibility among many due to the inherent ambiguity
introduced by regularization choices throughout the calculation process. Not
surprisingly, surface terms, rather than purely divergent terms, play the
primary role in this ambiguity. This finding underscores a fundamental
challenge: existing regularizations cannot simultaneously achieve both
mathematical consistency and physical meaning. It is not possible to find a
regularization capable of resolving the involved dilemma for the following reasons:

1) \textbf{Setting surface terms and their traces to zero:} While this
eliminates ambiguous terms often responsible for breaking symmetries, it does
so by violating the linearity of integration. This, in turn, leads to
non-unique results, undermining the ability to make genuine predictions.

2) \textbf{Retaining surface terms:} Although this preserves unique results
and avoids violating the linearity of integration, it also retains the
ambiguity and associated symmetry violations inherent in these terms.
Consequently, the resulting predictions remain ambiguous and lack clear
physical interpretation.

3) \textbf{Setting surface terms to zero but not their traces:} This approach
appears to offer a middle ground, but at the cost of violating the Schouten
identity. As before, this renders the results non-unique and non-predictive,
highlighting the impossibility of finding a "perfect" regularization that
satisfies both conditions.

In essence, regularizations introduce ambiguity, resulting in non-unique
outcomes. Furthermore, while surface terms hold the key to this ambiguity, no
existing regularization can simultaneously deliver both mathematically
consistent and physically meaningful results.

While the non-uniqueness and ambiguity caused by regularizations may be a
general concern in perturbative QFT calculations, it becomes particularly
critical in the context of anomalies. As demonstrated throughout this work,
achieving consistent results through regularizations in anomaly calculations
proves impossible across various examples. This necessitates exploring
alternative strategies beyond traditional Feynman rules to circumvent these
undesirable quantities and transform Feynman amplitudes into physically
meaningful ones. This transformation, akin to the removal of infinities in the
renormalization process without assuming them to be zero, requires a method
that does not rely on regularizations.

Finally, it is important to emphasize that investigation was made in a
completely regularization-free approach. Due to this it was possible to
appreciate some aspects in our analisys which are not possible to do in
contexts where regularizations are adopted.

\appendix

\section{Relations among Green functions \label{Ap_GFRs}}

The RAGFs involving the subamplitudes defined in (\ref{T2}), (\ref{T3}), and
(\ref{T4}) are presented in this appendix.

\subsection{Vector sector}

In the vector sector we have%
\begin{align}
p^{\sigma}\left[  T_{\mu;\sigma\rho}^{VV}\right]   &  =\left[  T_{\mu;\rho
}^{V}\left(  k_{1}\right)  \right]  -\left[  T_{\mu;\rho}^{V}\left(
k_{2}\right)  \right]  \ ,\label{RAGF_VV_3_index_p}\\
p^{\sigma}\left[  T_{\mu{\nu};\sigma\rho}^{VV}\right]   &  =\left[  T_{\mu
\nu;\rho}^{V}\left(  k_{1}\right)  \right]  -\left[  T_{\mu\nu;\rho}%
^{V}\left(  k_{2}\right)  \right]  \ , \label{RAGF_VV_4_index_p1}%
\end{align}%
\begin{align}
g^{\mu\sigma}\left[  T_{\mu;\sigma\rho}^{VV}\right]   &  =\left[  T_{\rho}%
^{V}\left(  k_{2}\right)  \right]  +m\left[  T_{\rho}^{SV}\right]
\ ,\label{RAGF_VV_3_index_g}\\
g^{\mu\sigma}\left[  T_{\mu{\nu};\sigma\rho}^{VV}\right]   &  =\left[
T_{\nu;\rho}^{V}\left(  k_{2}\right)  \right]  +m\left[  T_{\nu;\rho}%
^{SV}\right]  \ , \label{RAGF_VV_4_index_g}%
\end{align}%
\begin{align}
p^{\rho}\left[  T_{\rho;\mu\nu}^{VV}\right]   &  =-\frac{1}{2}g_{\mu\nu
}p^{\alpha}\left[  T_{\alpha}^{V}\left(  k_{1}\right)  +T_{\alpha}^{V}\left(
k_{2}\right)  \right] \nonumber\\
&  +\frac{1}{2}p_{\mu}\left[  T_{\nu}^{V}\left(  k_{1}\right)  -T_{\nu}%
^{V}\left(  k_{2}\right)  \right] \nonumber\\
&  +\frac{1}{2}p_{\nu}\left[  T_{\mu}^{V}\left(  k_{1}\right)  +T_{\mu}%
^{V}\left(  k_{2}\right)  \right] \nonumber\\
&  +\left[  T_{\mu;\nu}^{V}\left(  k_{1}\right)  -T_{\mu;\nu}^{V}\left(
k_{2}\right)  \right] \nonumber\\
&  -\frac{1}{2}p^{2}\left[  T_{\mu\nu}^{VV}\left(  k_{1},k_{2}\right)
\right]  \ ,
\end{align}%
\begin{align}
p^{\mu}\left[  T_{\mu\nu;\sigma\rho}^{VV}\right]   &  =-\frac{1}{2}%
g_{\sigma\rho}p^{\xi}\left[  T_{\nu;\xi}^{V}\left(  k_{1}\right)  +T_{\nu;\xi
}^{V}\left(  k_{2}\right)  \right] \nonumber\\
&  +\frac{1}{2}p_{\sigma}\left[  T_{\nu;\rho}^{V}\left(  k_{1}\right)
-T_{\nu;\rho}^{V}\left(  k_{2}\right)  \right] \nonumber\\
&  +\frac{1}{2}p_{\rho}\left[  T_{\nu;\sigma}^{V}\left(  k_{1}\right)
+T_{\nu;\sigma}^{V}\left(  k_{2}\right)  \right] \nonumber\\
&  +\left[  T_{\nu\rho;\sigma}^{V}\left(  k_{1}\right)  \right]  -\left[
T_{\nu\sigma;\rho}^{V}\left(  k_{2}\right)  \right] \nonumber\\
&  -\frac{1}{2}\left(  k_{2}-k_{1}\right)  ^{2}\left[  T_{\nu;\sigma\rho}%
^{VV}\right]  \ . \label{RAGF_VV_4_index_p2}%
\end{align}
Substituting these RAGFs in (\ref{V_Sector}) we get%
\begin{align}
p^{\mu}\left[  T_{\mu\nu\rho\sigma}^{\left(  V\right)  }\right]   &  =4\left[
T_{\nu\sigma;\rho}^{V}\left(  k_{1}\right)  \right]  -8\left[  T_{\nu
\sigma;\rho}^{V}\left(  k_{2}\right)  \right] \nonumber\\
&  +4\left[  T_{\nu\rho;\sigma}^{V}\left(  k_{1}\right)  \right]  -8\left[
T_{\nu\rho;\sigma}^{V}\left(  k_{2}\right)  \right]  +8\left[  T_{\sigma
\rho;\nu}^{V}\left(  k_{1}\right)  \right] \nonumber\\
&  -2g_{\nu\rho}p^{\xi}\left[  T_{\sigma;\xi}^{V}\left(  k_{1}\right)
+T_{\sigma;\xi}^{V}\left(  k_{2}\right)  \right]  -2g_{\nu\sigma}p^{\xi
}\left[  T_{\rho;\xi}^{V}\left(  k_{1}\right)  +T_{\rho;\xi}^{V}\left(
k_{2}\right)  \right] \nonumber\\
&  +4p_{\nu}\left[  T_{\sigma;\rho}^{V}\left(  k_{1}\right)  -T_{\sigma;\rho
}^{V}\left(  k_{2}\right)  \right]  +4p_{\nu}\left[  T_{\rho;\sigma}%
^{V}\left(  k_{1}\right)  -T_{\rho;\sigma}^{V}\left(  k_{2}\right)  \right]
\nonumber\\
&  +2p_{\rho}\left[  T_{\nu;\sigma}^{V}\left(  k_{1}\right)  -T_{\nu;\sigma
}^{V}\left(  k_{2}\right)  \right]  +2p_{\sigma}\left[  T_{\nu;\rho}%
^{V}\left(  k_{1}\right)  -T_{\nu;\rho}^{V}\left(  k_{2}\right)  \right]
\nonumber\\
&  +4p_{\rho}\left[  T_{\sigma;\nu}^{V}\left(  k_{1}\right)  \right]
+4p_{\sigma}\left[  T_{\rho;\nu}^{V}\left(  k_{1}\right)  \right] \nonumber\\
&  -g_{\nu\rho}p_{\sigma}p^{\xi}\left[  T_{\xi}^{V}\left(  k_{1}\right)
+T_{\xi}^{V}\left(  k_{2}\right)  \right]  -g_{\nu\sigma}p_{\rho}p^{\xi
}\left[  T_{\xi}^{V}\left(  k_{1}\right)  +T_{\xi}^{V}\left(  k_{2}\right)
\right] \nonumber\\
&  +2p_{\sigma}p_{\nu}\left[  T_{\rho}^{V}\left(  k_{1}\right)  -T_{\rho}%
^{V}\left(  k_{2}\right)  \right]  +2p_{\rho}p_{\nu}\left[  T_{\sigma}%
^{V}\left(  k_{1}\right)  -T_{\sigma}^{V}\left(  k_{2}\right)  \right]
\nonumber\\
&  +2p_{\sigma}p_{\rho}\left[  T_{\nu}^{V}\left(  k_{1}\right)  +T_{\nu}%
^{V}\left(  k_{2}\right)  \right]  \ . \label{p_V}%
\end{align}

\subsection{Axial-Vector sector}

The expected RAGFs for the subamplitudes in the axial-vector sector are%
\begin{align}
p^{\rho}T_{\sigma\rho}^{AV}  &  =\left[  T_{\sigma}^{A}\left(  k_{1}\right)
\right]  -\left[  T_{\sigma}^{A}\left(  k_{2}\right)  \right]
\ ,\label{RAGF_AV_2_index_p1}\\
p^{\sigma}T_{\sigma\rho}^{AV}  &  =\left[  T_{\rho}^{A}\left(  k_{1}\right)
\right]  -\left[  T_{\rho}^{A}\left(  k_{2}\right)  \right]  +2m\left[
T_{\rho}^{PV}\right]  \ , \label{RAGF_AV_2_index_p2}%
\end{align}%
\begin{align}
p^{\sigma}\left[  T_{\mu;\sigma\rho}^{AV}\right]   &  =\left[  T_{\mu;\rho
}^{A}\left(  k_{1}\right)  \right]  -\left[  T_{\mu;\rho}^{A}\left(
k_{2}\right)  \right]  +2m\left[  T_{\mu;\rho}^{PV}\right]
\ ,\label{RAGF_AV_3_index_p1}\\
p^{\sigma}\left[  T_{\mu\nu;\sigma\rho}^{AV}\right]   &  =\left[  T_{\mu
\nu;\rho}^{A}\left(  k_{1}\right)  \right]  -\left[  T_{\mu\nu;\rho}%
^{A}\left(  k_{2}\right)  \right]  +2m\left[  T_{\mu\nu;\rho}^{PV}\right]
\ ,\nonumber
\end{align}%
\begin{align}
g^{\mu\sigma}\left[  T_{\mu;\sigma\rho}^{AV}\right]   &  =\left[  T_{\rho}%
^{A}\left(  k_{2}\right)  \right]  -m\left[  T_{\rho}^{PV}\right]  \ ,\\
g^{\mu\sigma}\left[  T_{\mu\nu;\sigma\rho}^{AV}\right]   &  =\left[
T_{\nu;\rho}^{A}\left(  k_{2}\right)  \right]  -m\left[  T_{\nu;\rho}%
^{PV}\right]  \ ,
\end{align}%
\begin{align}
p^{\mu}\left[  T_{\mu;\sigma\rho}^{AV}\right]   &  =\frac{1}{2}\varepsilon
_{\rho\xi}p^{\xi}\left[  T_{\sigma}^{V}\left(  k_{1}\right)  +T_{\sigma}%
^{V}\left(  k_{2}\right)  \right] \nonumber\\
&  +\frac{1}{2}p_{\rho}\left[  T_{\sigma}^{A}\left(  k_{1}\right)  -T_{\sigma
}^{A}\left(  k_{2}\right)  \right] \nonumber\\
&  +\left[  T_{\rho;\sigma}^{A}\left(  k_{1}\right)  \right]  -\left[
T_{\rho;\sigma}^{A}\left(  k_{2}\right)  \right] \nonumber\\
&  -\frac{1}{2}p^{2}\left[  T_{\sigma\rho}^{AV}\right]  \ ,
\label{RAGF_AV_3_index_p2}%
\end{align}%
\begin{align}
p^{\mu}\left[  T_{\mu\nu;\sigma\rho}^{AV}\right]   &  =-\frac{1}{2}%
\varepsilon_{\rho\xi}p^{\xi}\left[  T_{\nu;\sigma}^{V}\left(  k_{1}\right)
+T_{\nu;\sigma}^{V}\left(  k_{2}\right)  \right] \nonumber\\
&  +\frac{1}{2}p_{\rho}\left[  T_{\nu;\sigma}^{A}\left(  k_{1}\right)
-T_{\nu;\sigma}^{A}\left(  k_{2}\right)  \right] \nonumber\\
&  +\left[  T_{\rho\nu;\sigma}^{A}\left(  k_{1}\right)  \right]  -\left[
T_{\rho\nu;\sigma}^{A}\left(  k_{2}\right)  \right] \nonumber\\
&  -\frac{1}{2}p^{2}\left[  T_{\nu;\sigma\rho}^{AV}\right]  \ .
\end{align}
Adding them up to the Eq. (\ref{AV_Sector}) we get%
\begin{align}
p^{\mu}\left[  \pm T_{\mu\nu\rho\sigma}^{\left(  AV\right)  }\right]   &
=4\left[  T_{\nu\sigma;\rho}^{A}\left(  k_{1}\right)  -T_{\nu\sigma;\rho}%
^{A}\left(  k_{2}\right)  \right]  +4\left[  T_{\nu\rho;\sigma}^{A}\left(
k_{1}\right)  -T_{\nu\rho;\sigma}^{A}\left(  k_{2}\right)  \right] \nonumber\\
&  +8\left[  T_{\sigma\rho;\nu}^{A}\left(  k_{1}\right)  -T_{\sigma\rho;\nu
}^{A}\left(  k_{2}\right)  \right]  +4p_{\rho}\left[  T_{\sigma;\nu}%
^{A}\left(  k_{1}\right)  -T_{\sigma;\nu}^{A}\left(  k_{2}\right)  \right]
\nonumber\\
&  +2p_{\rho}\left[  T_{\nu;\sigma}^{A}\left(  k_{1}\right)  -T_{\nu;\sigma
}^{A}\left(  k_{2}\right)  \right]  +4p_{\sigma}\left[  T_{\rho;\nu}%
^{A}\left(  k_{1}\right)  -T_{\rho;\nu}^{A}\left(  k_{2}\right)  \right]
\nonumber\\
&  +2p_{\sigma}\left[  T_{\nu;\rho}^{A}\left(  k_{1}\right)  -T_{\nu;\rho}%
^{A}\left(  k_{2}\right)  \right]  +2p_{\nu}\left[  T_{\sigma;\rho}^{A}\left(
k_{1}\right)  -T_{\sigma;\rho}^{A}\left(  k_{2}\right)  \right] \nonumber\\
&  +2p_{\nu}\left[  T_{\rho;\sigma}^{A}\left(  k_{1}\right)  -T_{\rho;\sigma
}^{A}\left(  k_{2}\right)  \right]  +2p_{\sigma}p_{\rho}\left[  T_{\nu}%
^{A}\left(  k_{1}\right)  -T_{\nu}^{A}\left(  k_{2}\right)  \right]
\nonumber\\
&  +p_{\nu}p_{\sigma}\left[  T_{\rho}^{A}\left(  k_{1}\right)  -T_{\rho}%
^{A}\left(  k_{2}\right)  \right]  +p_{\nu}p_{\rho}\left[  T_{\sigma}%
^{A}\left(  k_{1}\right)  -T_{\sigma}^{A}\left(  k_{2}\right)  \right]
\nonumber\\
&  -2\varepsilon_{\rho\xi}p_{\xi}\left[  T_{\sigma;\nu}^{V}\left(
k_{1}\right)  +T_{\sigma;\nu}^{V}\left(  k_{2}\right)  \right]  -2\varepsilon
_{\sigma\xi}p_{\xi}\left[  T_{\rho;\nu}^{V}\left(  k_{1}\right)  +T_{\rho;\nu
}^{V}\left(  k_{2}\right)  \right] \nonumber\\
&  -\varepsilon_{\rho\xi}p_{\sigma}p_{\xi}\left[  T_{\nu}^{V}\left(
k_{1}\right)  +T_{\nu}^{V}\left(  k_{2}\right)  \right]  -\varepsilon
_{\sigma\xi}p_{\rho}p_{\xi}\left[  T_{\nu}^{V}\left(  k_{1}\right)  +T_{\nu
}^{V}\left(  k_{2}\right)  \right] \nonumber\\
&  +8m\left[  T_{\nu\sigma;\rho}^{PV}+T_{\nu\rho;\sigma}^{PV}\right]
+4mp_{\nu}\left[  T_{\sigma;\rho}^{PV}+T_{\rho;\sigma}^{PV}\right]
+4mp_{\sigma}\left[  T_{\nu;\rho}^{PV}\right] \nonumber\\
&  +4mp_{\rho}\left[  T_{\nu;\sigma}^{PV}\right]  +2mp_{\nu}p_{\sigma}\left[
T_{\rho}^{PV}\right]  +2mp_{\nu}p_{\rho}\left[  T_{\sigma}^{PV}\right]  \ ,
\end{align}%
\begin{align}
p^{\mu}\left[  \pm T_{\mu\nu\rho\sigma}^{\left(  VA\right)  }\right]   &
=8\left[  T_{{\nu\sigma;}\rho}^{A}\left(  k_{1}\right)  -T_{{\nu\sigma;}\rho
}^{A}\left(  k_{2}\right)  \right]  +8\left[  T_{{\nu\rho;\sigma}}^{A}\left(
k_{1}\right)  -T_{{\nu\rho;\sigma}}^{A}\left(  k_{2}\right)  \right]
\nonumber\\
&  +4p_{\nu}\left[  T_{\sigma;\rho}^{A}\left(  k_{1}\right)  -T_{\sigma;\rho
}^{A}\left(  k_{2}\right)  +T_{\rho;\sigma}^{A}\left(  k_{1}\right)
-T_{\rho;\sigma}^{A}\left(  k_{2}\right)  \right] \nonumber\\
&  +4p_{\sigma}\left[  T_{\nu;\rho}^{A}\left(  k_{1}\right)  -T_{\nu;\rho}%
^{A}\left(  k_{2}\right)  \right]  +4p_{\rho}\left[  T_{\nu;\sigma}^{A}\left(
k_{1}\right)  -T_{\nu;\sigma}^{A}\left(  k_{2}\right)  \right] \nonumber\\
&  +2p_{\sigma}p_{\nu}\left[  T_{\rho}^{A}\left(  k_{1}\right)  -T_{\rho}%
^{A}\left(  k_{2}\right)  \right]  +2p_{\rho}p_{\nu}\left[  T_{\sigma}%
^{A}\left(  k_{1}\right)  -T_{\sigma}^{A}\left(  k_{2}\right)  \right]
\nonumber\\
&  -2\varepsilon_{\nu\xi}p^{\xi}\left[  T_{\sigma;\rho}^{V}\left(
k_{1}\right)  +T_{\sigma;\rho}^{V}\left(  k_{2}\right)  +T_{\rho;\sigma}%
^{V}\left(  k_{1}\right)  +T_{\rho;\sigma}^{V}\left(  k_{2}\right)  \right]
\nonumber\\
&  -\varepsilon_{\nu\xi}p^{\xi}\left[  p_{\sigma}T_{\rho}^{V}\left(
k_{1}\right)  +p_{\sigma}T_{\rho}^{V}\left(  k_{2}\right)  +p_{\rho}T_{\sigma
}^{V}\left(  k_{1}\right)  +p_{\rho}T_{\sigma}^{V}\left(  k_{2}\right)
\right]  \ . \label{p_VA}%
\end{align}

\subsection{Axial sector \label{Ap_RAGF_Axial_Sec}}

The expected RAGFs for the subamplitudes in the axial sector are%
\begin{equation}
p^{\sigma}\left[  T_{\sigma\rho}^{AA}\right]  =\left[  T_{\rho}^{V}\left(
k_{1}\right)  \right]  -\left[  T_{\rho}^{V}\left(  k_{2}\right)  \right]
+2m\left[  T_{\rho}^{PA}\right]  \ ,
\end{equation}%
\begin{equation}
p^{\sigma}\left[  T_{\mu;\sigma\rho}^{AA}\right]  =\left[  T_{\mu;\rho}%
^{V}\left(  k_{1}\right)  \right]  -\left[  T_{\mu;\rho}^{V}\left(
k_{2}\right)  \right]  +2m\left[  T_{\mu;\rho}^{PA}\right]  \ ,
\end{equation}%
\begin{equation}
p^{\sigma}\left[  T_{\mu\nu;\sigma\rho}^{AA}\right]  =\left[  T_{\mu\nu;\rho
}^{V}\left(  k_{1}\right)  \right]  -\left[  T_{\mu\nu;\rho}^{V}\left(
k_{2}\right)  \right]  +2m\left[  T_{\mu\nu;\rho}^{PA}\right]  \ ,
\end{equation}%
\begin{equation}
g^{\mu\sigma}\left[  T_{\mu;\sigma\rho}^{AA}\right]  =\left[  T_{\rho}%
^{V}\left(  k_{2}\right)  \right]  -m\left[  T_{\rho}^{PA}\right]  \ ,
\end{equation}%
\begin{equation}
g^{\mu\sigma}\left[  T_{\mu\nu;\sigma\rho}^{AA}\right]  =\left[  T_{\nu;\rho
}^{V}\left(  k_{2}\right)  \right]  -m\left[  T_{\nu;\rho}^{PA}\right]  \ ,
\label{RAGF_AA_4_index_g}%
\end{equation}%
\begin{align}
p^{\mu}\left[  T_{\mu;\sigma\rho}^{AA}\right]   &  =-\frac{1}{2}g_{\sigma\rho
}p^{\xi}\left[  T_{\xi}^{V}\left(  k_{1}\right)  +T_{\xi}^{V}\left(
k_{2}\right)  \right] \nonumber\\
&  +\frac{1}{2}p_{\sigma}\left[  T_{\rho}^{V}\left(  k_{1}\right)  -T_{\rho
}^{V}\left(  k_{2}\right)  \right] \nonumber\\
&  +\frac{1}{2}p_{\rho}\left[  T_{\sigma}^{V}\left(  k_{1}\right)  +T_{\sigma
}^{V}\left(  k_{2}\right)  \right] \nonumber\\
&  +\left[  T_{\rho;\sigma}^{V}\left(  k_{1}\right)  \right]  -\left[
T_{\sigma;\rho}^{V}\left(  k_{2}\right)  \right] \nonumber\\
&  -mg_{\sigma\rho}\left[  T^{S}\left(  k_{1}\right)  -T^{S}\left(
k_{2}\right)  \right] \nonumber\\
&  -\frac{1}{2}p^{2}\left[  T_{\sigma\rho}^{AA}\right]  \ ,
\end{align}%
\begin{align}
p^{\mu}\left[  T_{\mu\nu;\sigma\rho}^{AA}\right]   &  =-\frac{1}{2}%
g_{\sigma\rho}p^{\xi}\left[  T_{\nu;\xi}^{V}\left(  k_{1}\right)  +T_{\nu;\xi
}^{V}\left(  k_{2}\right)  \right] \nonumber\\
&  +\frac{1}{2}p_{\sigma}\left[  T_{\nu;\rho}^{V}\left(  k_{1}\right)
-T_{\nu;\rho}^{V}\left(  k_{2}\right)  \right] \nonumber\\
&  +\frac{1}{2}p_{\rho}\left[  T_{\nu;\sigma}^{V}\left(  k_{1}\right)
+T_{\nu;\sigma}^{V}\left(  k_{2}\right)  \right] \nonumber\\
&  +\left[  T_{\nu\rho;\sigma}^{V}\left(  k_{1}\right)  \right]  -\left[
T_{\nu\sigma;\rho}^{V}\left(  k_{2}\right)  \right] \nonumber\\
&  -mg_{\sigma\rho}\left[  T_{\nu}^{S}\left(  k_{1}\right)  -T_{\nu}%
^{S}\left(  k_{2}\right)  \right] \nonumber\\
&  -\frac{1}{2}p^{2}\left[  T_{\nu;\sigma\rho}^{AA}\right]  \ .
\end{align}
Replacing them into (\ref{A_Sector}) gives%
\begin{align}
p^{\mu}\left[  T_{\mu\nu\rho\sigma}^{\left(  A\right)  }\right]   &  =4\left[
T_{\nu\sigma;\rho}^{V}\left(  k_{1}\right)  \right]  -8\left[  T_{\nu
\sigma;\rho}^{V}\left(  k_{2}\right)  \right]  +4\left[  T_{\nu\rho;\sigma
}^{V}\left(  k_{1}\right)  \right]  -8\left[  T_{\nu\rho;\sigma}^{V}\left(
k_{2}\right)  \right] \nonumber\\
&  +8\left[  T_{\sigma\rho;\nu}^{V}\left(  k_{1}\right)  \right]  -2g_{\nu
\rho}p_{\xi}\left[  T_{\sigma;\xi}^{V}\left(  k_{1}\right)  +T_{\sigma;\xi
}^{V}\left(  k_{2}\right)  \right] \nonumber\\
&  -2g_{\nu\sigma}p_{\xi}\left[  T_{\rho;\xi}^{V}\left(  k_{1}\right)
+T_{\rho;\xi}^{V}\left(  k_{2}\right)  \right] \nonumber\\
&  +4p_{\nu}\left[  T_{\sigma;\rho}^{V}\left(  k_{1}\right)  -T_{\sigma;\rho
}^{V}\left(  k_{2}\right)  +T_{\rho;\sigma}^{V}\left(  k_{1}\right)
-T_{\rho;\sigma}^{V}\left(  k_{2}\right)  \right] \nonumber\\
&  +2p_{\rho}\left[  T_{\sigma;\nu}^{V}\left(  k_{1}\right)  +T_{\sigma;\nu
}^{V}\left(  k_{2}\right)  \right]  +2p_{\rho}\left[  T_{\sigma;\nu}%
^{V}\left(  k_{1}\right)  +T_{\nu;\sigma}^{V}\left(  k_{1}\right)  \right]
\nonumber\\
&  +2p_{\sigma}\left[  T_{\rho;\nu}^{V}\left(  k_{1}\right)  +T_{\rho;\nu}%
^{V}\left(  k_{2}\right)  \right]  +2p_{\sigma}\left[  T_{\rho;\nu}^{V}\left(
k_{1}\right)  +T_{\nu;\rho}^{V}\left(  k_{1}\right)  \right] \nonumber\\
&  -4p_{\sigma}\left[  T_{\nu;\rho}^{V}\left(  k_{2}\right)  \right]
-4p_{\rho}\left[  T_{\nu;\sigma}^{V}\left(  k_{2}\right)  \right] \nonumber\\
&  -\left(  g_{\nu\rho}p_{\sigma}+g_{\nu\sigma}p_{\rho}\right)  p^{\xi}\left[
T_{\xi}^{V}\left(  k_{1}\right)  +T_{\xi}^{V}\left(  k_{2}\right)  \right]
\nonumber\\
&  +2p_{\nu}p_{\sigma}\left[  T_{\rho}^{V}\left(  k_{1}\right)  -T_{\rho}%
^{V}\left(  k_{2}\right)  \right]  +2p_{\nu}p_{\rho}\left[  T_{\sigma}%
^{V}\left(  k_{1}\right)  -T_{\sigma}^{V}\left(  k_{2}\right)  \right]
\nonumber\\
&  +2p_{\sigma}p_{\rho}\left[  T_{\nu}^{V}\left(  k_{1}\right)  +T_{\nu}%
^{V}\left(  k_{2}\right)  \right] \nonumber\\
&  -4mg_{\nu\rho}\left[  T_{\sigma}^{S}\left(  k_{1}\right)  -T_{\sigma}%
^{S}\left(  k_{2}\right)  \right]  -2mg_{\nu\rho}p_{\sigma}\left[
T^{S}\left(  k_{1}\right)  -T^{S}\left(  k_{2}\right)  \right] \nonumber\\
&  -4mg_{\nu\sigma}\left[  T_{\rho}^{S}\left(  k_{1}\right)  -T_{\rho}%
^{S}\left(  k_{2}\right)  \right]  -2mg_{\nu\sigma}p_{\rho}\left[
T^{S}\left(  k_{1}\right)  -T^{S}\left(  k_{2}\right)  \right] \nonumber\\
&  +8m\left[  T_{\nu\sigma;\rho}^{PA}+T_{\nu\rho;\sigma}^{PA}\right]
+4mp_{\nu}\left[  T_{\sigma;\rho}^{PA}+T_{\rho;\sigma}^{PA}\right]
+4mp_{\sigma}\left[  T_{\nu;\rho}^{PA}\right] \nonumber\\
&  +4mp_{\rho}\left[  T_{\nu;\sigma}^{PA}\right]  +2mp_{\nu}p_{\sigma}\left[
T_{\rho}^{PA}\right]  +2mp_{\nu}p_{\rho}\left[  T_{\sigma}^{PA}\right]  \ .
\label{p_A}%
\end{align}

\section{Integrals Results}

In order to perform the required calculations to obtain the subamplitudes
shown above, it is enough to use the integrals results which we list in the
following. They are
\begin{equation}
I_{2}=\frac{i}{\left(  4\pi\right)  }\left[  \xi_{0}^{\left(  -1\right)
}\left(  p^{2},m^{2}\right)  \right]  \ ,
\end{equation}%
\begin{equation}
I_{2\mu}=-\frac{i}{\left(  4\pi\right)  }p_{\mu}\left[  \xi_{1}^{\left(
-1\right)  }\left(  p^{2},m^{2}\right)  \right]  +\left(  -k_{1}\right)
_{\mu}\left[  I_{2}\right]  \ ,
\end{equation}%
\begin{align}
I_{2\mu\nu}  &  =\frac{1}{2}\left[  \Delta_{2;\mu\nu}^{\left(  2\right)
}\left(  \lambda^{2}\right)  \right]  +\frac{1}{2}g_{\mu\nu}\left[  I_{\log
}^{\left(  2\right)  }\left(  \lambda^{2}\right)  \right] \nonumber\\
&  -\frac{i}{\left(  4\pi\right)  }\frac{1}{2}g_{\mu\nu}\left[  \xi
_{0}^{\left(  0\right)  }\left(  p^{2},m^{2};\lambda^{2}\right)  \right]
\nonumber\\
&  +\frac{i}{\left(  4\pi\right)  }p_{\mu}p_{\nu}\left[  \xi_{2}^{\left(
-1\right)  }\left(  p^{2},m^{2}\right)  \right] \nonumber\\
&  +\left(  -k_{1}\right)  _{\mu}\left[  I_{2\nu}\right]  +\left(
-k_{1}\right)  _{\nu}\left[  I_{2\mu}\right] \nonumber\\
&  -\left(  k_{1}\right)  _{\nu}\left(  k_{1}\right)  _{\mu}\left[
I_{2}\right]  \ ,
\end{align}

\begin{align}
I_{2\mu\nu\lambda}  &  =-\frac{1}{4}\left(  k_{2}+k_{1}\right)  ^{\xi}\left[
\square_{3;\xi\mu{{\nu}\lambda}}^{\left(  2\right)  }\left(  \lambda
^{2}\right)  \right] \nonumber\\
&  -\frac{1}{8}\left(  k_{2}+k_{1}\right)  ^{\xi}\left\{  g_{\mu{{\nu}}%
}\left[  \Delta_{2;\lambda\xi}^{\left(  2\right)  }\left(  \lambda^{2}\right)
\right]  +g_{\mu{\lambda}}\left[  \Delta_{2;{\nu}\xi}^{\left(  2\right)
}\left(  \lambda^{2}\right)  \right]  +g_{{\nu}\lambda}\left[  \Delta
_{2;\mu\xi}^{\left(  2\right)  }\left(  \lambda^{2}\right)  \right]  \right\}
\nonumber\\
&  -\frac{1}{8}\left(  k_{2}+k_{1}\right)  _{\mu}\left[  \Delta_{2;{\nu
}\lambda}^{\left(  2\right)  }\left(  \lambda^{2}\right)  \right]  -\frac
{1}{8}\left(  k_{2}+k_{1}\right)  _{{\nu}}\left[  \Delta_{2;\mu{\lambda}%
}^{\left(  2\right)  }\left(  \lambda^{2}\right)  \right]  -\frac{1}{8}\left(
k_{2}+k_{1}\right)  _{\lambda}\left[  \Delta_{2;\mu{{\nu}}}^{\left(  2\right)
}\left(  \lambda^{2}\right)  \right] \nonumber\\
&  +\frac{1}{2}\left(  k_{1}\right)  _{\mu}\left[  \Delta_{2;\nu\lambda
}^{\left(  2\right)  }\left(  \lambda^{2}\right)  \right]  +\frac{1}{2}\left(
k_{1}\right)  _{\nu}\left[  \Delta_{2;\mu\lambda}^{\left(  2\right)  }\left(
\lambda^{2}\right)  \right]  +\frac{1}{2}\left(  k_{1}\right)  _{\lambda
}\left[  \Delta_{2;\mu\nu}^{\left(  2\right)  }\left(  \lambda^{2}\right)
\right] \nonumber\\
&  -\frac{1}{4}\left[  g_{\mu{{\nu}}}\left(  k_{2}-k_{1}\right)  _{\lambda
}+g_{\mu{\lambda}}\left(  k_{2}-k_{1}\right)  _{{{\nu}}}+g_{{{\nu}\lambda}%
}\left(  k_{2}-k_{1}\right)  _{{\mu}}\right]  \left[  I_{\log}^{\left(
2\right)  }\left(  \lambda^{2}\right)  \right] \nonumber\\
&  +\frac{1}{2}g_{\mu\nu}p_{\lambda}\left[  \xi_{1}^{\left(  0\right)
}\left(  p^{2};m^{2};\lambda^{2}\right)  \right]  +\frac{1}{2}g_{\mu\lambda
}p_{\nu}\left[  \xi_{1}^{\left(  0\right)  }\left(  p^{2};m^{2};\lambda
^{2}\right)  \right] \nonumber\\
&  +\frac{1}{2}g_{\nu\lambda}p_{\mu}\left[  \xi_{1}^{\left(  0\right)
}\left(  p^{2};m^{2};\lambda^{2}\right)  \right]  -p_{\mu}p_{\nu}p_{\lambda
}\left[  \xi_{3}^{\left(  -1\right)  }\left(  p^{2};m^{2}\right)  \right]
\nonumber\\
&  +\left(  -k_{1}\right)  _{\mu}\left[  I_{2\nu\lambda}\right]  +\left(
-k_{1}\right)  _{\nu}\left[  I_{2\mu\lambda}\right]  +\left(  -k_{1}\right)
_{\lambda}\left[  I_{2\mu\nu}\right] \nonumber\\
&  -\left(  -k_{1}\right)  _{\mu}\left(  -k_{1}\right)  _{\nu}\left[
I_{2\lambda}\right]  -\left(  -k_{1}\right)  _{\mu}\left(  -k_{1}\right)
_{\lambda}\left[  I_{2\nu}\right]  -\left(  -k_{1}\right)  _{\nu}\left(
-k_{1}\right)  _{\lambda}\left[  I_{2\mu}\right] \nonumber\\
&  +\left(  -k_{1}\right)  _{{\mu}}\left(  -k_{1}\right)  _{\nu}\left(
-k_{1}\right)  _{{\lambda}}\left[  I_{2}\right]  \ .
\end{align}

\section{One-point Functions}

The one-point functions which are used in this work are defined by%
\begin{equation}
T_{\sigma}^{i}\left(  k_{1}\right)  =\int\frac{d^{2}k}{\left(  2\pi\right)
^{2}}Tr\left\{  \left[  \Gamma_{i}\right]  _{\sigma}\frac{1}{\not k  +\not k
_{1}-m}\right\}  ~,
\end{equation}%
\begin{equation}
T_{\mu;\sigma}^{i}\left(  k_{1}\right)  =\int\frac{d^{2}k}{\left(
2\pi\right)  ^{2}}\left(  k+k_{1}\right)  _{\mu}Tr\left\{  \left[  \Gamma
_{i}\right]  _{\sigma}\frac{1}{\not k  +\not k  _{1}-m}\right\}  \ ,
\end{equation}

\begin{equation}
T_{\mu\nu;\sigma}^{i}\left(  k_{1}\right)  =\int\frac{d^{2}k}{\left(
2\pi\right)  ^{2}}\left(  k+k_{1}\right)  _{\mu}\left(  k+k_{1}\right)  _{\nu
}Tr\left\{  \left[  \Gamma_{i}\right]  _{\sigma}\frac{1}{\not k  +\not k
_{1}-m}\right\}  \ ,
\end{equation}
where $\Gamma_{i}$ are vertex operators belonging to the set $\Gamma
_{i}=\left\{  \Gamma_{S},\Gamma_{P},\Gamma_{V},\Gamma_{A}\right\}
=\{1,\gamma_{3},\gamma_{\alpha},\gamma_{\alpha}\gamma_{3}\}$. By using the
integrals shown above one obtain the following results,%
\begin{equation}
T^{S}\left(  k_{1}\right)  =2m\left[  I_{\log}^{\left(  2\right)  }\left(
m^{2}\right)  \right]  \ ,
\end{equation}%
\begin{equation}
T_{\nu}^{S}\left(  k_{2}\right)  =2m\left\{  -k_{2}^{\xi}\left[  \Delta
_{2;\nu\xi}^{\left(  2\right)  }\left(  m^{2}\right)  \right]  -p_{\nu}\left[
I_{\log}^{\left(  2\right)  }\left(  m^{2}\right)  \right]  \right\}  \ ,
\end{equation}%
\begin{align}
T_{\sigma}^{A}  &  =-\varepsilon_{\sigma\xi}g^{\xi\chi}T_{\chi}^{V}%
\ ,\label{One_point_A_1_index}\\
T_{\mu;\sigma}^{A}  &  =-\varepsilon_{\sigma\xi}g^{\xi\chi}T_{\mu;\chi}%
^{V}\ ,\\
T_{\mu\nu;\sigma}^{A}  &  =-\varepsilon_{\sigma\xi}g^{\xi\chi}T_{\mu\nu;\chi
}^{V}\ ,
\end{align}%
\begin{equation}
T_{\mu}^{V}\left(  k_{i}\right)  =-2k_{i}^{\xi}\left[  \Delta_{2;\mu\xi
}^{\left(  2\right)  }\right]  \ , \label{One_point_V_1_index}%
\end{equation}%
\begin{align}
T_{\nu;\mu}^{V}\left(  k_{2}\right)   &  =g_{\mu\nu}\left[  I_{quad}^{\left(
2\right)  }\left(  m^{2}\right)  \right]  +\left[  \Delta_{1;\mu\nu}^{\left(
2\right)  }\left(  m^{2}\right)  \right] \nonumber\\
&  +k_{2}^{\xi}k_{2}^{\chi}\left\{  \left[  \Delta_{3;\mu\nu\xi\chi}^{\left(
2\right)  }\left(  m^{2}\right)  \right]  +\frac{1}{2}g_{\mu\nu}\left[
\Delta_{2;\xi\chi}^{\left(  2\right)  }\left(  m^{2}\right)  \right]  \right\}
\nonumber\\
&  -\frac{1}{2}k_{2}^{2}\left[  \Delta_{2;\mu\nu}^{\left(  2\right)  }\left(
m^{2}\right)  \right]  -k_{2\mu}k_{2}^{\xi}\left[  \Delta_{2;\nu\xi}^{\left(
2\right)  }\left(  m^{2}\right)  \right] \nonumber\\
&  +\left(  k_{2\nu}k_{2}^{\xi}-2k_{1\nu}k_{2}^{\xi}\right)  \left[
\Delta_{2;\mu\xi}^{\left(  2\right)  }\left(  m^{2}\right)  \right]  \ ,
\label{One_point_V_2_index}%
\end{align}%
\begin{align}
T_{\mu\nu;\sigma}^{V}\left(  k_{2}\right)   &  =\left[  \left(  k_{1}%
-k_{2}\right)  _{\mu}g_{\nu\sigma}+\left(  k_{1}-k_{2}\right)  _{\nu}%
g_{\mu\sigma}\right]  \left[  I_{quad}^{\left(  2\right)  }\left(
m^{2}\right)  \right]  -\frac{1}{3}k_{2}^{\xi}k_{2}^{\chi}k_{2}^{\omega
}\left[  \Sigma_{4;\mu\nu\sigma\omega\xi\chi}^{\left(  2\right)  }\right]
\nonumber\\
&  -k_{2}^{\xi}\left[  \square_{2;\mu\nu\sigma\xi}^{\left(  2\right)  }\left(
m^{2}\right)  \right]  -\left(  k_{2}-k_{1}\right)  _{\mu}k_{2}^{\xi}%
k_{2}^{\chi}\left[  \square_{3;\nu\sigma\xi\chi}^{\left(  2\right)  }\left(
m^{2}\right)  \right]  +k_{2}^{2}k_{2}^{\xi}\left[  \square_{3;\mu\nu\sigma
\xi}^{\left(  2\right)  }\left(  m^{2}\right)  \right] \nonumber\\
&  +k_{2}^{\xi}k_{2}^{\chi}\left\{  k_{2\sigma}\left[  \square_{3;\mu\nu
\xi\chi}^{\left(  2\right)  }\left(  m^{2}\right)  \right]  +k_{1\nu}\left[
\square_{3;\mu\sigma\xi\chi}^{\left(  2\right)  }\left(  m^{2}\right)
\right]  \right. \nonumber\\
&  \ \ \ \ \ \ \ \ \ \ \ \ \ \left.  -\frac{1}{3}g_{\mu\nu}k_{2}^{\omega
}\left[  \square_{3;\sigma\omega\chi\xi}^{\left(  2\right)  }\left(
m^{2}\right)  \right]  -\frac{1}{3}g_{\mu\sigma}k_{2}^{\omega}\left[
\square_{3;\nu\omega\xi\chi}^{\left(  2\right)  }\left(  m^{2}\right)
\right]  \right\} \nonumber\\
&  +k_{2\sigma}\left[  \Delta_{1;\mu\nu}^{\left(  2\right)  }\left(
m^{2}\right)  \right]  +k_{1\nu}\left[  \Delta_{1;\mu\sigma}^{\left(
2\right)  }\left(  m^{2}\right)  \right]  -\left(  k_{2}-k_{1}\right)  _{\mu
}\left[  \Delta_{1;\nu\sigma}^{\left(  2\right)  }\left(  m^{2}\right)
\right] \nonumber\\
&  -k_{2}^{\xi}\left\{  g_{\mu\nu}\left[  \Delta_{1;\sigma\xi}^{\left(
2\right)  }\left(  m^{2}\right)  \right]  +g_{\mu\sigma}\left[  \Delta
_{1;\nu\xi}^{\left(  2\right)  }\left(  m^{2}\right)  \right]  \right\}
\nonumber\\
&  -k_{2}^{2}k_{2\sigma}\left[  \Delta_{2;\mu\nu}^{\left(  2\right)  }\left(
m^{2}\right)  \right]  -k_{2}^{2}k_{1\mu}\left[  \Delta_{2;\nu\sigma}^{\left(
2\right)  }\left(  m^{2}\right)  \right]  +k_{2}^{2}\left(  k_{2}%
-k_{1}\right)  _{\nu}\left[  \Delta_{2;\mu\sigma}^{\left(  2\right)  }\left(
m^{2}\right)  \right] \nonumber\\
&  -2k_{1\mu}k_{2\sigma}k_{2}^{\xi}\left[  \Delta_{2;\nu\xi}^{\left(
2\right)  }\left(  m^{2}\right)  \right]  +\left[  2\left(  k_{2}%
-k_{1}\right)  _{\nu}k_{2\sigma}+k_{2}^{2}g_{\nu\sigma}\right]  k_{2}^{\xi
}\left[  \Delta_{2;\mu\xi}^{\left(  2\right)  }\left(  m^{2}\right)  \right]
\nonumber\\
&  +\left[  k_{2}^{2}g_{\mu\nu}-2\left(  k_{2}-k_{1}\right)  _{\mu}\left(
k_{2}-k_{1}\right)  _{\nu}\right]  k_{2}^{\xi}\left[  \Delta_{2;\sigma\xi
}^{\left(  2\right)  }\left(  m^{2}\right)  \right] \nonumber\\
&  -\left[  g_{\mu\sigma}\left(  k_{2}-k_{1}\right)  _{\nu}+g_{\nu\sigma
}\left(  k_{2}-k_{1}\right)  _{\mu}\right]  k_{2}^{\xi}k_{2}^{\chi}\left[
\Delta_{2;\xi\chi}^{\left(  2\right)  }\left(  m^{2}\right)  \right]  \ .
\end{align}

\section{Two-point functions \label{Ap_2P_Func}}

Here we show some results for the two-point functions not shown explicitly in
the main text. They are:%

\begin{align}
S_{{\mu\nu\sigma\rho}}^{VV}  &  =\left[  \square_{2;\mu\nu\sigma\rho}^{\left(
2\right)  }\left(  m^{2}\right)  \right]  +\frac{1}{12}\left(  3P^{\xi}%
P^{\chi}+p^{\xi}p^{\chi}\right)  \left[  \Sigma_{4;\mu\nu\sigma\rho\xi\chi
}^{\left(  2\right)  }\left(  m^{2}\right)  \right]  -\frac{1}{4}\left(
P^{2}+p^{2}\right)  \left[  \square_{3;\mu\nu\sigma\rho}^{\left(  2\right)
}\left(  m^{2}\right)  \right] \nonumber\\
&  -\frac{1}{2}\left(  P_{\nu}-p_{\nu}\right)  P^{\xi}\left[  \square
_{3;\mu\sigma\rho\xi}^{\left(  2\right)  }\left(  m^{2}\right)  \right]
-\frac{1}{2}P_{{\sigma}}P^{\xi}\left[  \square_{3;\mu\nu\rho\xi}^{\left(
2\right)  }\left(  m^{2}\right)  \right]  -\frac{1}{2}P_{{\rho}}P^{\xi}\left[
\square_{3;\mu\nu\sigma\xi}^{\left(  2\right)  }\left(  m^{2}\right)  \right]
\nonumber\\
&  +\frac{1}{12}\left(  P_{\mu}P^{\xi}+p_{\mu}p^{\xi}-p_{\mu}P^{\xi}+P_{\mu
}p^{\xi}\right)  \left[  \square_{3;\nu\sigma\rho\chi}^{\left(  2\right)
}\left(  m^{2}\right)  \right]  -\frac{1}{4}g_{{\sigma\rho}}\left(  P^{\xi
}P^{\chi}+p^{\xi}p^{\chi}\right)  \left[  \square_{3;\mu\nu\xi\chi}^{\left(
2\right)  }\left(  m^{2}\right)  \right] \nonumber\\
&  +\frac{1}{12}\left(  3P^{\xi}P^{\chi}+p^{\xi}p^{\chi}\right)  \left\{
g_{\mu\nu}\left[  \square_{3;\sigma\rho\chi\xi}^{\left(  2\right)  }\left(
m^{2}\right)  \right]  +g_{\mu\sigma}\left[  \square_{3;\nu\rho\xi\chi
}^{\left(  2\right)  }\left(  m^{2}\right)  \right]  +g_{\mu\rho}\left[
\square_{3;\nu\sigma\xi\chi}^{\left(  2\right)  }\left(  m^{2}\right)
\right]  \right\} \nonumber\\
&  -g_{{\sigma\rho}}\left[  \Delta_{1;\mu\nu}^{\left(  2\right)  }\left(
m^{2}\right)  \right]  +g_{\mu\nu}\left[  \Delta_{1;\sigma\rho}^{\left(
2\right)  }\left(  m^{2}\right)  \right]  +g_{\mu\sigma}\left[  \Delta
_{1;\nu\rho}^{\left(  2\right)  }\left(  m^{2}\right)  \right]  +g_{\mu\rho
}\left[  \Delta_{1;\nu\sigma}^{\left(  2\right)  }\left(  m^{2}\right)
\right] \nonumber\\
&  +\left[  \frac{1}{6}p_{\mu}p_{\nu}-\frac{1}{4}g_{\mu\nu}\left(  P^{2}%
+p^{2}\right)  \right]  \left[  \Delta_{2;\sigma\rho}^{\left(  2\right)
}\left(  m^{2}\right)  \right]  +\frac{1}{2}g_{{\sigma\rho}}\left(  P_{\nu
}-p_{\nu}\right)  P^{\xi}\left[  \Delta_{2;\mu\xi}^{\left(  2\right)  }\left(
m^{2}\right)  \right] \nonumber\\
&  +\frac{1}{2}\left[  g_{{\sigma\rho}}\left(  P^{2}+3p^{2}\right)
+P_{\sigma}P_{\rho}-p_{\sigma}p_{\rho}\right]  \left[  \Delta_{2;\mu\nu
}^{\left(  2\right)  }\left(  m^{2}\right)  \right] \nonumber\\
&  -\frac{1}{4}g_{\mu\sigma}\left(  P^{2}+p^{2}\right)  \left[  \Delta
_{2;\nu\rho}^{\left(  2\right)  }\left(  m^{2}\right)  \right]  -\frac{1}%
{4}g_{\mu\rho}\left(  P^{2}+p^{2}\right)  \left[  \Delta_{2;\nu\sigma
}^{\left(  2\right)  }\left(  m^{2}\right)  \right] \nonumber\\
&  +\frac{1}{2}\left(  P_{\nu}-p_{\nu}\right)  P_{{\sigma}}\left[
\Delta_{2;\mu\rho}^{\left(  2\right)  }\left(  m^{2}\right)  \right]
+\frac{1}{2}\left(  P_{\nu}-p_{\nu}\right)  P_{{\rho}}\left[  \Delta
_{2;\mu\sigma}^{\left(  2\right)  }\left(  m^{2}\right)  \right] \nonumber\\
&  -\frac{1}{2}\left[  g_{\mu\sigma}P_{{\rho}}P^{\xi}+g_{\mu\rho}P_{{\sigma}%
}P^{\xi}-g_{\sigma\rho}P_{\mu}P^{\xi}+g_{{\sigma\rho}}\left(  P^{\xi}P_{\mu
}+p^{\xi}p_{\mu}\right)  \right]  \left[  \Delta_{2;\nu\xi}^{\left(  2\right)
}\left(  m^{2}\right)  \right] \nonumber\\
&  -\frac{1}{2}\left[  g_{\mu\nu}P_{{\rho}}P^{\xi}-g_{\mu\rho}\left(
-\frac{1}{2}P^{\xi}P_{\nu}+p_{\nu}P^{\xi}+\frac{1}{6}p^{\xi}p_{\nu}\right)
-\frac{1}{3}g_{\nu\rho}p^{\xi}p_{\mu}\right]  \left[  \Delta_{2;\sigma\xi
}^{\left(  2\right)  }\left(  m^{2}\right)  \right] \nonumber\\
&  +\frac{1}{2}\left[  \frac{1}{3}g_{\nu\sigma}p^{\xi}p_{\mu}-\frac{1}%
{6}g_{\mu\nu}\left(  3P^{\xi}P_{\sigma}-p^{\xi}p_{\sigma}\right)
+g_{\mu\sigma}\left(  \frac{1}{3}p^{\xi}p_{\nu}+p_{\nu}P^{\xi}\right)
\right]  \left[  \Delta_{2;\rho\xi}^{\left(  2\right)  }\left(  m^{2}\right)
\right] \nonumber\\
&  -\frac{1}{6}\left[  g_{\mu\nu}g_{\sigma\rho}p^{\xi}p^{\chi}-\frac{1}%
{2}\left(  g_{\mu\sigma}g_{\nu\rho}+g_{\mu\rho}g_{\nu\sigma}\right)  \left(
3P^{\xi}P^{\chi}+p^{\xi}p^{\chi}\right)  \right]  \left[  \Delta_{2;\chi\xi
}^{\left(  2\right)  }\left(  m^{2}\right)  \right]  \ ,
\end{align}

\begin{equation}
T_{\sigma}^{SV}=0\ , \label{two_point_SV_1_index}%
\end{equation}%
\begin{align}
T_{\nu;\sigma}^{SV}  &  =2m\left\{  \left[  \Delta_{2;\nu\sigma}^{\left(
2\right)  }\left(  m^{2}\right)  \right]  +g_{\nu\sigma}\left[  I_{\log
}^{\left(  2\right)  }\left(  m^{2}\right)  \right]  \right\} \nonumber\\
&  -2m\left\{  g_{\nu\sigma}\left[  \xi_{0}^{\left(  0\right)  }\left(
p^{2},m^{2};m^{2}\right)  \right]  -p_{\nu}p_{\sigma}\left[  2\xi_{2}^{\left(
-1\right)  }\left(  p^{2},m^{2}\right)  -\xi_{1}^{\left(  -1\right)  }\left(
p^{2},m^{2}\right)  \right]  \right\}  \ , \label{two_point_SV_2_index}%
\end{align}%
\begin{equation}
T_{\sigma}^{PV}=\frac{i}{2\pi}m\varepsilon_{\sigma\xi}p^{\xi}\left[  \xi
_{0}^{\left(  -1\right)  }\left(  p^{2},m^{2}\right)  \right]  ~,
\label{two_point_PV_1_index}%
\end{equation}%
\begin{equation}
T_{\sigma}^{PA}=-2mp_{\sigma}\left[  \xi_{0}^{\left(  -1\right)  }\left(
p^{2},m^{2}\right)  \right]  \ ,
\end{equation}%
\begin{equation}
T_{\nu;\sigma}^{PV}=-\frac{1}{2}p_{\nu}\left[  T_{\sigma}^{PV}\right]  \ ,
\end{equation}%
\begin{equation}
T_{\mu;\sigma}^{PA}=-\frac{1}{2}p_{\mu}\left[  T_{\sigma}^{PA}\right]  \ ,
\end{equation}%
\begin{align}
T_{\mu\nu;\sigma}^{PV}  &  =m\varepsilon_{\sigma\xi}p^{\xi}\left\{  \left[
\Delta_{2;\mu\nu}^{\left(  2\right)  }\left(  m^{2}\right)  \right]
+g_{\mu\nu}\left[  I_{\log}^{\left(  2\right)  }\left(  m^{2}\right)  \right]
\right\} \nonumber\\
&  +m\varepsilon_{\sigma\xi}p^{\xi}\left(  p_{\mu}p_{\nu}-g_{\mu\nu}%
p^{2}\right)  \left[  2\xi_{2}^{\left(  -1\right)  }\left(  p^{2}%
,m^{2}\right)  -\xi_{1}^{\left(  -1\right)  }\left(  p^{2},m^{2}\right)
\right] \nonumber\\
&  +m\varepsilon_{\sigma\xi}p^{\xi}p_{\mu}p_{\nu}\left[  \xi_{1}^{\left(
-1\right)  }\left(  p^{2},m^{2}\right)  \right]  \ ,
\end{align}%
\begin{align}
T_{\mu\nu;\sigma}^{PA}  &  =-mp_{\sigma}\left\{  \left[  \Delta_{2;\mu\nu
}^{\left(  2\right)  }\left(  m^{2}\right)  \right]  +g_{\mu\nu}\left[
I_{\log}^{\left(  2\right)  }\left(  m^{2}\right)  \right]  \right\}
\nonumber\\
&  +mp_{\sigma}\left\{  \left(  p_{\mu}p_{\nu}-g_{\mu\nu}p^{2}\right)  \left[
\xi_{1}^{\left(  -1\right)  }\left(  p^{2},m^{2}\right)  -2\xi_{2}^{\left(
-1\right)  }\left(  p^{2},m^{2}\right)  \right]  \right. \nonumber\\
&  \ \ \ \ \ \ \ \ \ \ \ \left.  -p_{\mu}p_{\nu}\left[  \xi_{1}^{\left(
-1\right)  }\left(  p^{2},m^{2}\right)  \right]  \right\}  \ .
\end{align}

\end{document}